\documentclass[11pt,letterpaper]{article}
\usepackage[T1]{fontenc}
\usepackage{lmodern}
\usepackage[margin=1in]{geometry}
\usepackage{amsmath,amssymb,amsthm,mathtools}
\usepackage{booktabs,array}
\usepackage{graphicx}
\usepackage{pgfplots}
\pgfplotsset{compat=1.18}
\usepackage{microtype}
\usepackage{enumitem}
\usepackage[numbers,sort&compress]{natbib}
\usepackage{xcolor}
\definecolor{inkblue}{HTML}{244A68}
\usepackage[colorlinks=true,linkcolor=inkblue,citecolor=inkblue,urlcolor=inkblue]{hyperref}
\hypersetup{pdftitle={The Theory of Strategic Evolution: Three Principles of Strategic Replicators},pdfauthor={Kevin Vallier}}
\setlist[enumerate]{itemsep=0.2em,topsep=0.4em}
\newtheorem{theorem}{Theorem}[section]
\newtheorem{proposition}[theorem]{Proposition}

\theoremstyle{definition}
\newtheorem{definition}[theorem]{Definition}
\newcommand{\doi}[1]{\href{https://doi.org/#1}{\nolinkurl{doi:#1}}}
\title{The Theory of Strategic Evolution:\\[0.25em]
Three Principles of Strategic Replicators}
\author{Kevin Vallier\\[0.4em]
\small Institute of American Constitutional Thought and Leadership\\
\small University of Toledo}
\date{Working draft --- October 13, 2026}
\begin{document}
\maketitle
\begin{abstract}
Strategic replicators combine two mechanisms usually distinguished in theoretical explanation: evaluated choice and reproductive transmission. An agent chooses among alternatives; the consequences of those choices influence continuation or reproduction; and successors receive organization that shapes later choices. Artificial intelligence makes this combination economically salient when agents have authority to create, configure, or retire software instances. This paper organizes the general logic around three principles: situated choice, reproductive transmission, and institution-mediated coupling. The principles are constitutive rather than predictions of growth or improvement. A common representation distinguishes carriers, inherited records, decision owners, and the mechanisms that allocate consequences. Actual parent--descendant attribution yields a selection--transmission identity. Separate compatibility results characterize when utility-maximizing choices also maximize a reproductive criterion. Gode and Sunder's zero-intelligence traders reveal the role of institutions: enforced properties can hold across different decision procedures, while aggregate efficiency alone determines neither individual gains nor descent. A worked market--rent--entry model connects institutional profits to actual offspring, with its occurrence-level ancestry bridge checked in Lean. A direct-deployment example then shows evaluated preferences undergoing selection without presupposing that every chooser values reproduction. Formal appendices develop games with endogenous players, deployment-induced tipping, endogenous electorates, and the production of institutions, with explicit models, proofs, and worked examples. The paper and its supporting formal artifacts are a prototype kernel with a generative function: an instrument for constructing models of strategic replication. The results are formal; their application to deployed populations and their empirical implications remain to be tested.
\end{abstract}
\noindent\textbf{Keywords:} strategic replicators; evolutionary theory; choice; institutions; artificial intelligence; zero-intelligence traders.
\clearpage

\section*{A note on method}

My findings are formal. The paper develops a general account of strategic
replication and proves results about specified models. Selected finite
foundations are also machine-checked in Lean. These are not yet empirical
findings about populations of deployed AI systems. A proof establishes what
follows from a model's premises; it does not establish that a particular
population satisfies those premises.

The paper and its supporting formal artifacts should therefore be understood
as a \emph{prototype kernel with a generative function}: an instrument for
studying the social world of strategic replicators. The Three Principles
articulate the combined logic of choice and reproductive transmission.
Adding assumptions about technologies, information, objectives, resources,
and institutions produces more specific models, including market tipping,
price formation, and elections. Several such models have been developed in
companion papers and are brought together in the appendices. Their empirical
implications still require testing. Testing may require substantial
simulation resources, access to deployment data, or both. The original
project did not have the resources to undertake that empirical program,
and no empirical validation is claimed here. The generality of the principles serves their
foundational purpose, while a particular model must earn its explanatory
and predictive application to the world.

AI is the motivating application, but the account is substrate-neutral.
It concerns systems that realize the relations among choice, transmission,
and institutional consequences specified below. Strategic replication does
not require superintelligence. Applying the account to a particular
technology nevertheless requires establishing the relevant capacities,
permissions, and transmission mechanisms.

The framework, the interpretation, and the applications are mine. The
mathematical exposition was developed with AI assistance under my direction.
The original paper documented a Lean repository as its primary formal
artifact. This expanded account combines that formal program with
analytical proofs; Appendix~\ref{app:formal-scope} states exactly which
results have been machine-checked. The supporting code is an inspectable
part of the research instrument, and its verification claims apply to the
encoded statements under their hypotheses.

I chose the collaborative method deliberately. When the project began,
self-replication as a strategic variable had little readership, and I
expected much of the work to reach readers through language models. I
therefore wrote it to be legible to those models as well as to human readers.
In this methodological sense, the paper is an instance of its own thesis:
a human-directed collective of human and artificial reasoners producing a
result through their combined work. This collaboration is part of the
work's provenance; it is not evidence for the empirical adequacy of its
models.

\clearpage

\section{Introduction}
\label{sec:introduction}

Societies change when their capital changes. The transition from agriculture to industry transformed social relationships and political institutions. Artificial intelligence raises a further possibility: productive assets can participate in decisions about their own deployment. A software system may select tasks, hire services, request additional instances, or alter the procedures those instances execute. Whether it can do any of these things depends on its permissions, resources, and surrounding institutions. When such capabilities are available, the population of decision-making systems becomes one of the consequences of their decisions.

The resulting question is conceptual before it is predictive. What is the general logic of a system that both chooses and reproduces? Economics studies how alternatives are selected in light of evaluations and constraints. Evolutionary theory studies how differences in contribution and transmission change the representation of traits. Each logic is useful at a level of generality that precedes a particular prediction. A choice model needs preferences and feasible alternatives before it predicts demand. A selection model needs a reproductive process and transmission assumptions before it predicts the evolution of a trait. The corresponding task for strategic replicators is to explain how these mechanisms fit together.

Consider an AI controller choosing between completing a task itself and deploying two successors. The first option may conserve resources, reduce coordination costs, and preserve control. The second may expand productive capacity while introducing competition, correlated errors, or new decision owners. The controller evaluates these alternatives; an institutional and physical mechanism determines which requests can be executed; and the resulting successors inherit some organization from the system that produced them. Their subsequent choices affect later reproduction. Choice and selection concern the same continuing process.

This paper develops the Theory of Strategic Evolution around three principles: \emph{situated choice}, \emph{reproductive transmission}, and \emph{institution-mediated coupling}. A strategic replicator is an organized system in which evaluated choices can affect its continuation or reproduction, and reproductive transmission can carry decision-relevant organization into future choosers. In the principal economic case, the system also has authority over a deployment channel: it can request the creation, retirement, or configuration of instances. This direct authority is a substantive feature of a model, not something supplied by the fact that software is technically copyable.

The distinction between a lineage and an instance is especially useful here. A continuing controller can choose a deployment architecture and the number of instances operating within it. Some instances are transient workers controlled by a continuing owner. Others become decision owners themselves. Shared code or ancestry does not settle which description is appropriate. The theory therefore distinguishes the carrier that exists, the organization that is transmitted, and the owner who chooses.

The intellectual motivation brings together two parts of von Neumann's research: strategic choice and the construction of reproducing machines \cite{vonNeumannMorgenstern1944,vonNeumann1966}. Preference evolution, evolutionary games, and controlled branching already connect important parts of the problem. The contribution proposed here is an explicit organization of the combined role, together with mathematical interfaces connecting choice, institutional consequences, and actual descent. The elementary accounting identities used below make those interfaces precise.

Gode and Sunder's zero-intelligence traders provide a crucial comparison \cite{GodeSunder1993}. Their results direct attention to the division of explanatory labor between individual procedures and institutional rules. A rule can exclude loss-making trades even when a trader generates random bids. A favorable aggregate outcome therefore need not reveal an optimizing individual. Conversely, a rational individual need not improve aggregate welfare. Adding reproduction introduces a further question: which consequences accrue to which carriers, and what do those carriers transmit? Market efficiency alone does not answer it.

The paper develops three connections in detail. First, an actual parent--descendant map yields a selection/transmission identity. This is a statement about represented occurrences and traits, with no presumption that a mean increases. Second, a choice optimum becomes a reproductive optimum only under a compatibility condition linking the two orderings. Third, an institution's guarantees transfer across policies that respect its enforced conditions, while distribution-dependent performance need not transfer. A worked market-and-reproduction model then connects attributed profits to survival, spawning, and the resulting selection identity. The occurrence-level connection has been checked in Lean.

The level of abstraction is deliberate. A general principle can delimit a mechanism without selecting a unique future trajectory. Predictive models arise by specifying information, objectives, inheritance, resource flows, and interaction rules. The foundational question is which assumptions belong to the combined role and which belong to a particular result. Separating them preserves the possibility of a theory that expands in applications while retaining a small common structure.

The discussion of AI alignment follows this distinction. Stability, welfare, safe operation, and alignment are different properties. Preferences influence actions, institutions influence their consequences, and reproduction influences which procedures remain represented. Their joint study is necessary for population-level analysis, but the three principles do not establish a general tendency toward either alignment or its failure. The final sections identify the additional questions a substantive application must answer.

\paragraph{Method and scope.}
This is a foundational and mathematical paper. The examples are specified models, and the new formal verification checks exact finite statements. They do not constitute experimental evidence about deployed AI populations. The paper distinguishes constitutive principles, mathematical deductions, and optional model assumptions. A formalization appendix records the scope of the machine-checked results; the proofs in the text are intended to stand independently of the software implementation.


\section{Choice, selection, and the construction of successors}
\label{sec:literature-lineages}

Strategic replication connects several established questions. Decision and game theory ask how agents evaluate alternatives when their consequences depend on uncertainty and other agents. Evolutionary theory asks how differential contribution and transmission change a population. The theory of self-reproducing automata asks how organized systems can construct successors. A general account must connect these questions without treating their answers as interchangeable.

Von Neumann and Morgenstern's \emph{Theory of Games and Economic Behavior} supplies a foundational treatment of strategic interaction \cite{vonNeumannMorgenstern1944}. Von Neumann's later work on self-reproducing automata, edited and completed by Burks, investigates reproduction through specified construction mechanisms \cite{vonNeumann1966}. For present purposes, the distinction is between evaluating an alternative and implementing the mechanism that produces its consequences. An instruction to create a copy does not determine whether creation is feasible, what is copied, or who controls the resulting system. Conversely, a mechanism capable of copying organization need not evaluate whether it ought to do so. These become connected when the objects being reproduced include procedures for making subsequent choices.

Price's covariance formulations provide a general language for tracking differential contribution and changes during transmission \cite{Price1970,Price1972}. They do not require a population to behave like a single optimizing individual. The selection identity is informative precisely because it separates components of change without assuming that every component has the same sign. Frank emphasizes the difference between this abstract accounting and a dynamically sufficient model \cite{Frank2012}. We use that distinction constructively: a common representation identifies what models share, while additional mechanisms determine which trajectories occur.

The connection between choice and selection is also established within economics. In the indirect evolutionary approach, preferences influence behavior while a possibly different material criterion determines evolutionary success. Dekel, Ely, and Yilankaya study this relationship in games where the observability of preferences affects interaction \cite{DekelElyYilankaya2007}. Their framework is a direct antecedent to the distinction between an evaluator and induced reproductive consequences. The questions pursued here concern the transmitted organization that implements choice and the authority agents have over creating, configuring, and controlling successors.

Grafen's Formal Darwinism program offers another relevant comparison. It connects population-genetic change with individual optimization through an explicit mapping from genotype to phenotype and reproductive consequences \cite{Grafen2008}. The optimization of an individual's phenotype is distinguished from maximizing a function of population frequencies. This supports a methodological choice made here: the relation between individual evaluation and population change should be supplied by a bridge with stated assumptions, rather than asserted through an undifferentiated principle of improvement.

Nor does a changing number of agents place a model outside existing game theory. Claisse, Ren, and Tan develop mean-field games with branching, allowing the population to change while agents control their dynamics \cite{ClaisseRenTan2023}. Their distinction between individual agents' objectives and control of an entire branching population is especially relevant. A parent controlling several workers and a parent producing autonomous successors are different strategic structures, even when the physical offspring counts agree. The three principles organize such differences; they do not replace the analytical tools or existence hypotheses of branching-game theory.

The resulting claim is one of explicit organization and connection. We distinguish the carrier that is reproduced, the record that is transmitted, the owner who chooses, and the institution that allocates consequences. These distinctions permit familiar models to be recovered while showing where further assumptions enter.

\section{Three principles of strategic replication}
\label{sec:principles}

The proposed principles identify a combined role. They specify what makes a process one of strategic replication, leaving particular theories of rationality, growth, and equilibrium to add their own assumptions. Their usefulness lies in distinguishing the mechanisms that must be connected and the conclusions that require something further.

\subsection{Situated choice}

\paragraph{Principle 1.} \emph{A strategic agent uses an evaluative procedure to choose among alternatives available under its information and authority.}

Evaluation gives alternatives significance for the chooser. It can concern consumption, task completion, the continuation of a project, descendants, or the conduct of future agents. The procedure may be bounded, fallible, or sensitive to several considerations that do not admit a single numerical ranking. The principle requires an interpreted connection between evaluation and choice; an arbitrary constant utility attached retrospectively to a random sequence would not establish that connection.

Expected-utility maximization supplies an important specialization. Let $\Pi_o(h)$ be the admissible policies for decision owner $o$ at history $h$, let $U_o$ evaluate complete consequences, and let $\mu_o$ describe that owner's beliefs. Exact optimization requires
\[
 \pi_o\in\mathop{\rm arg\,max}_{\pi\in\Pi_o(h)}
       \mathbb E_{\mu_o,\pi}[U_o],
\]
when this maximum exists. The alternatives, beliefs, and treatment of ties are part of the specification. Maximizing a current reward is a different restriction from maximizing the value of a complete policy. Neither restriction by itself supplies correct beliefs, common knowledge, or equilibrium among several owners.

An owner's information constrains the decision problem as well as the selected action. If two histories are indistinguishable to that owner, a model cannot silently make different hidden facts available through their menus. Where a request may fail because of an unknown resource constraint, the local alternative can be to submit the request; the execution mechanism then determines whether it succeeds. Authority likewise concerns permissions. A program that can describe a copy instruction does not thereby possess the credentials, resources, or rights needed to execute it.

The common mathematical setting can also contain procedures that do not satisfy this principle. A zero-intelligence trader can generate random quotes within an enforced range. It need not evaluate the quotes as a utility maximizer. A principal might deliberately select that random procedure as a cheap trading instrument, but the principal's evaluated choice and the instrument's random choices are different explanatory claims. Keeping both in one model permits a comparison between institutions and policies without defining every policy as rational agency.

\subsection{Reproductive transmission}

\paragraph{Principle 2.} \emph{A causal reproduction mechanism can produce distinct successor carriers and transmit organization relevant to their subsequent choices.}

A carrier is an occurrence of the relevant organized system: for example, a running program instance or an embodied organism. Its transmitted organization might include an evaluator, a decision procedure, a program description, or memory. The model must specify what is copied, what is transformed, and how the resulting record enters later decisions. A trait shared by parent and child is not, by itself, evidence that the parent supplied it. Both might instead acquire the trait from a common environment.

Replication requires a distinct successor, but it need not increase population size. A parent can produce one replacement and disappear. Conversely, a carrier can persist, learn, or revise its procedure without reproducing. A population-accounting convention may count a surviving parent among its own successors; that convention does not turn ordinary persistence into reproduction. Distinguishing the physical event from the accounting convention allows the same selection identity to handle survival and births without conflating them.

Inheritance need not preserve everything. An agent may copy its task procedure while changing its successor's objective or deleting its memory. Faithful transmission of a lineage label does not by itself establish faithful transmission of a policy, and faithful transmission of a policy does not imply repetition of an action: inherited procedures can respond differently to different observations. Changes in transmitted organization therefore belong inside the theory, with exact copying treated as a useful special case.

Ancestry also differs from authority. A child might remain an instrument of its parent's controller, acquire an independent controller, or pass to another owner. None of these arrangements follows merely from descent. Likewise, several carriers can share a controller without sharing an ancestor. The appropriate units of choice and selection must be declared separately before their relationship can be studied.

\subsection{Institution-mediated coupling}

\paragraph{Principle 3.} \emph{Feasible choices can affect continuation or reproduction through the environment and its institutions, while transmitted organization can affect successors' choices within the same process.}

This principle excludes placing an unrelated optimizer beside an unrelated copying process and calling the conjunction a synthesis. The connection has two directions. Different admissible choices can alter an observable of the continuing or descendant population---its size, resources, organization, or future rights. Differences in transmitted organization can alter a successor's choice disposition. A concrete model should display the relevant counterfactual comparisons, holding fixed the background conditions that define each comparison. Observational correlation alone does not establish either channel.

Neither channel must be active at every history. All currently represented agents might choose the same action, inherit the same procedure, or leave the same number of successors. Differential selection can then vanish even though reproductive and evaluative capacities remain. The principles describe a structured capability, not a requirement that every round exhibit growth, competition, or change. Feedback from population composition to payoffs can be important, but frequency dependence is an additional feature, not a universal requirement.

Institutions specify permissions, interaction, attribution, and execution. In an economic model they can include trading rules, property claims, prices, credit arrangements, and access to compute. Physical resource limits also enter the consequence mechanism. No literal market or institutional designer is required for the general principle. Where agents can amend rules, the amendment procedure and the authority to invoke it are themselves part of the represented system.

These distinctions explain why intelligence, profitability, and reproduction cannot be identified. A successful trade can increase resources without authorizing a copy. An authorized copy can consume resources without being profitable. A profitable carrier can transmit a procedure that its controller would prefer to eliminate. The causal account must state which consequences accrue to which carriers and how those consequences affect continuation and transmission. No common scalar is required to represent utility, money, descendants, and welfare.

\section{A common mathematical representation}
\label{sec:representation}

The three principles admit several mathematical realizations. A history-based representation is convenient because it does not require population shares, a fixed list of owners, or a stationary environment to be sufficient state variables. The finite accounting results below need much less structure than a complete stochastic process.

\subsection{Carriers, records, and decision owners}

At time $t$, let $C_t$ be the set of carrier occurrences and $O_t$ the set of decision owners. The sets may change. An occurrence $i\in C_t$ has a record $\theta_{t,i}$ containing the organization relevant to the modeled decisions: for example, a program, evaluator, memory state, or deployment policy. A separate ownership relation records who may make which decisions for that occurrence. One owner may control many carriers; a carrier can have different authorities for different actions. Ownership is not inferred from ancestry.

Write $h_t$ for the relevant history. It includes carrier records, resources, interactions, institutional state, and any chance variables the model needs. An owner $o$ observes $I_o(h_t)$ and chooses a local request from a menu $A_o(I_o(h_t))$. A policy has the form
\begin{equation}
  \pi_o(\,\cdot\mid I_o(h_t),\theta_o),
  \label{eq:local-policy}
\end{equation}
where $\theta_o$ denotes the decision-relevant organization available to that owner. The notation does not assume that all owners have numerical utility functions or that their policies are independent. Private randomization, common shocks, communication, and public correlation can be included explicitly.

A local request need not be globally feasible. Two owners can each request the last available unit of compute. The interaction mechanism resolves the conflict, records the rejection or allocation, and specifies what the parties learn. Defining a local menu using hidden global state without specifying how that information is revealed would silently change the decision problem. Where feasibility is not locally observable, it is better to model requests and enforcement separately.

\subsection{Consequences and transmission}

Let $a_t$ denote the joint requests. An interaction mechanism $M$ maps $h_t$, $a_t$, and chance input $\xi_t$ to an intermediate consequence record $y_t$. This record includes allocations, payments, physical events, and the attribution of consequences. A continuation mechanism $R$ maps the resulting situation to successor occurrences, transmitted records, and future rights. Schematically,
\begin{equation}
 (h_t,a_t,\xi_t)\xrightarrow{M}y_t,
 \qquad
 (h_t,a_t,y_t,\eta_t)\xrightarrow{R}
 (C_{t+1},\theta_{t+1},O_{t+1},h_{t+1}).
 \label{eq:mechanism}
\end{equation}
The second chance input may be dependent on the first; a joint law must be supplied if randomness matters. Factoring a mechanism this way is a way of stating its meaning, not an assumption of statistical independence or an assertion that reproduction always happens in a physically separate phase.

For a finite, closed, single-parent realization, $R$ also supplies an attribution map
\begin{equation}
 p_t:C_{t+1}\longrightarrow C_t.
 \label{eq:parent-map}
\end{equation}
A persisting parent can count as its own successor. A new child is attributed to the carrier that produced it. The map records actual occurrences rather than an assignment obtained by comparing their final traits. An external entrant has no parent in $C_t$; it requires an entry term or an enlarged attribution domain. Merging and component-wise inheritance can require weighted or multiple-parent attribution. These possibilities fit the broader principles but are outside the single-parent specialization used for the elementary theorem.

Two meanings of transmission should be distinguished. The mechanism can establish a causal channel by which organization is copied or transformed. Separately, a particular measured trait may be faithful, in the sense that its successor value equals its assigned parent's value. Causal copying can occur with changes in the trait, and equal trait values can occur without causal copying. The covariance identity concerns the specified attribution and trait values; it does not identify the copying mechanism from observational resemblance.

\subsection{What the representation does and does not compress}

The joint action law induced by the policies, combined with $M$ and $R$, determines a transition law when the required probability structure is supplied. An arbitrary transition law, however, does not by itself identify which variables represent evaluation, authority, inheritance, or enforcement. The three principles concern that interpreted structure. Their nondegeneracy requirement excludes a choice process and a reproducing process that merely happen to be written in the same state vector.

Further simplification requires further assumptions. A recursion on type shares is closed only if omitted resources, memories, information, and response choices do not matter beyond those shares. A fixed reproduction matrix can be useful, but it may hide the very strategic and institutional dependence at issue. An aggregate population statistic can be sufficient for one observable and insufficient for another. For example, total surplus determines neither its distribution across traders nor the number of children produced by each lineage.

The representation measure is also a modeling choice. Instance count, represented compute, capital, and beneficial ownership need not change together. Splitting one worker into two can double instance count while leaving total compute and ownership unchanged. The results below use occurrence count and say so explicitly. Capital-weighted selection or component inheritance requires an appropriate contribution measure rather than an unannounced change in the meaning of fitness.

No unlimited-time existence result is implied by this notation. For an infinite stochastic system one must supply measurable spaces, admissible policies, a well-defined transition law, and conditions preventing pathological accumulation of events where needed. The foundational principles allow such models; they do not discharge their analytic hypotheses. The finite realization makes the core connection available without first imposing a particular continuous-time growth law.

\section{Reproductive accounting and the transmission of choice}
\label{math-accounting}

The Three Principles identify a kind of system. To describe its evolution,
we must additionally specify what is represented in a population and how that
representation passes between censuses. Instance count, capital and computing
capacity are different measures. A reproduction event can increase one while
conserving another. The following result begins with actual finite censuses,
so that its denominator and attribution are explicit.

Let $I$ be a set of $n>0$ current occurrences and $J$ a set of $m>0$
successor occurrences. A specified map $p:J\to I$ assigns each successor to
one current parent. A persistent carrier may count as its own continuation.
Define
\[
 w_i=\#\{j\in J:p(j)=i\},\qquad
 \bar w=\frac1n\sum_{i\in I}w_i=\frac mn.
\]
Occurrences remain distinct even when their records are identical. The map is
part of the account of reproduction; resemblance alone does not determine it.
For real observables $z_i$ and $z'_j$, write
\[
 \bar z=\frac1n\sum_i z_i,\qquad
 \bar z'=\frac1m\sum_j z'_j,\qquad
 \operatorname{Cov}_I(w,z)=\frac1n\sum_i w_i z_i-\bar w\bar z.
\]

\begin{theorem}[Selection and transmission]
\label{math-price}
For these actual censuses and their parent map,
\begin{equation}
 \bar w(\bar z'-\bar z)
 =\operatorname{Cov}_I(w,z)
   +\frac1n\sum_{j\in J}\bigl(z'_j-z_{p(j)}\bigr).
 \label{math-price-equation}
\end{equation}
\end{theorem}

\begin{proof}
Grouping successors by their assigned parents gives
$\sum_j z_{p(j)}=\sum_i w_i z_i$. Consequently the right-hand side is
\[
 \frac1n\sum_i w_i z_i-\frac mn\bar z
 +\frac1n\sum_j z'_j-\frac1n\sum_i w_i z_i
 =\frac mn(\bar z'-\bar z).
\]
\end{proof}

This is the finite selection--transmission accounting associated with the
Price equation \cite{Price1970,Price1972}. The first term records differential
representation of current traits; the second records their alteration in
transmission. It neither requires maximizing agents nor proves improvement.
If all contributions are equal, the covariance vanishes, but transmission
can still change the population. If every successor copies its parent's
trait, the transmission term vanishes, but the covariance can have either
sign. Extinction is a legitimate outcome; when $m=0$, there is no normalized
successor mean to insert into this equation.

The same algebra permits other measures and declared fractional attribution.
For normalized nonnegative current weights $\rho_i$ and nonnegative contributions
$M_{ij}$, put $w_i=\sum_jM_{ij}$, $W=\sum_i \rho_iw_i>0$ and
$\rho'_j=\sum_i \rho_iM_{ij}/W$. Then
\begin{equation}
 W\left(\sum_j\rho'_jz'_j-\sum_i\rho_i z_i\right)
 =\operatorname{Cov}_{\rho}(w,z)
 +\sum_i \rho_i\left(\sum_jM_{ij}z'_j-w_i z_i\right).
 \label{math-weighted-price}
\end{equation}
An attribution for multiple parents must be declared rather than inferred
from this identity. External immigration likewise needs an additional
contribution term. In stochastic models, realized counts have random
denominators: normalizing expected counts generally gives a different
quantity from the expected normalized population.

\subsection{Inherited dispositions and newly chosen acts}
\label{math-dispositions}

Decision-relevant organization can be inherited without the parent's last
action being copied. Let $G_i$ be parent $i$'s standardized mean action
statistic in a specified decision context. Conditional on the realized
censuses and parent map, suppose a finitely supported joint law of fresh successor actions
$A'_j$ satisfies
\begin{equation}
 \mathbb E[g(A'_j)\mid I,J,p]=G_{p(j)}
 \quad\text{for every }j.
 \label{math-reset}
\end{equation}
The conditioning also fixes the context and parental records defining $G$.
The actions may share random draws and be arbitrarily correlated.
Write $\bar g'=m^{-1}\sum_jg(A'_j)$. Linearity of expectation and Theorem~\ref{math-price} yield
\begin{equation}
 \frac mn\left(
 \mathbb E[\bar g'\mid I,J,p]
 -\frac1n\sum_iG_i\right)
 =\operatorname{Cov}_I(w,G).
 \label{math-action-price}
\end{equation}
Indeed, the conditional successor mean is
$m^{-1}\sum_jG_{p(j)}$, to which the faithful-trait identity applies.

Condition~\eqref{math-reset} is substantive. Copying code does not establish
it when descendants encounter different information, resources or contexts.
Conditioning on survival or the realized offspring census can itself change
action distributions. Nor may $G_i$ silently be replaced by the parent's
realized action statistic: that comparison retains a transmission term.
The result concerns fresh choices conditional on actual descent, and does
not identify selection of a procedure with repetition of a fortunate act.

\section{When preferred choices also favor reproduction}
\label{math-compatibility}

The first principle permits evaluated choice; the third connects its
consequences to reproduction. Their conjunction does not identify the two
criteria. At a fixed decision situation, let $A$ be a nonempty finite set of
complete alternatives. Let $u:A\to\mathbb R$ represent the controller's
evaluation and let $r:A\to\mathbb R$ be a separately specified reproductive
criterion, such as expected descendant count over a declared horizon.
For a nonempty menu $B\subseteq A$, write $\operatorname{argmax}_B u$ for
the set of its $u$-maximizers.

\begin{theorem}[Compatibility on all menus]
\label{math-menus}
The following conditions are equivalent:
\begin{enumerate}
 \item For every nonempty menu $B\subseteq A$,
 $\operatorname{argmax}_B u\subseteq\operatorname{argmax}_B r$.
 \item For all $a,b\in A$, $u(a)\ge u(b)$ implies $r(a)\ge r(b)$.
\end{enumerate}
When they hold, $r=f\circ u$ on $A$ for a nondecreasing function $f$
defined on the attained utility values.
\end{theorem}

\begin{proof}
Suppose the first condition holds and $u(a)\ge u(b)$. Then $a$ maximizes
$u$ on $\{a,b\}$, and therefore maximizes $r$ there; hence $r(a)\ge r(b)$.
Conversely, if $a$ maximizes $u$ on $B$, applying the second condition to
each $b\in B$ shows that it maximizes $r$ there. Applying the order condition
in both directions shows that $u(a)=u(b)$ entails $r(a)=r(b)$. Thus
$f(u(a))=r(a)$ is well-defined, and the same condition makes $f$
nondecreasing.
\end{proof}

The requirement concerns every maximizing choice, so utility ties must be
reproductive ties. A particular tie-breaking rule or a restricted family of
menus can support weaker conditions. Optimization alone supplies none of
them: with $u(a)=1>u(b)=0$ and $r(a)=1<r(b)=2$, the uniquely preferred
alternative reproduces less. The theorem identifies the additional bridge
needed for compatibility; it does not make compatibility constitutive of
agency.

\begin{theorem}[Compatibility over lotteries]
\label{math-lotteries}
Let $X$ be a nonempty finite outcome space with real scores $u(x)$ and $r(x)$.
Evaluate every lottery $q\in\Delta(X)$ by
$U(q)=\sum_xq_xu(x)$ and $R(q)=\sum_xq_xr(x)$.
Every $U$-maximizer in every nonempty finite menu of lotteries also maximizes
$R$ if and only if
\begin{equation}
 r(x)=\alpha u(x)+\beta\quad\text{for all }x\in X,
 \qquad \alpha\ge0.
 \label{math-affine}
\end{equation}
\end{theorem}

\begin{proof}
An affine relation with nonnegative slope gives
$R(q)=\alpha U(q)+\beta$, which preserves maximizing choices. For necessity,
two-lottery menus show that equal $U$ entails equal $R$, and greater $U$
cannot entail smaller $R$. If $u$ is constant, comparison of degenerate
lotteries makes $r$ constant too; choose $\alpha=0$.

Otherwise choose outcomes $x_-,x_+$ attaining distinct minimum and maximum
utilities $u_-,u_+$. Set
$\alpha=(r(x_+)-r(x_-))/(u_+-u_-)\ge0$ and
$\beta=r(x_-)-\alpha u_-$. For any $x$, let
$\lambda=(u(x)-u_-)/(u_+-u_-)\in[0,1]$.
The degenerate lottery at $x$ and the mixture assigning probabilities
$1-\lambda,\lambda$ to $x_-,x_+$ have equal expected utility. They therefore
have equal expected reproductive score, giving
$r(x)=(1-\lambda)r(x_-)+\lambda r(x_+)=\alpha u(x)+\beta$.
\end{proof}

The full lottery domain makes this stronger than deterministic order
agreement. For example, squaring the scores $0,1,2$ preserves their order
but assigns different reproductive scores to the middle outcome and the
equally useful half-and-half mixture of the extremes. Constant reproductive
scores remain compatible; nonconstant $r$ requires $\alpha>0$.
The theorem assumes expected-score evaluation on a common outcome space.
It does not derive that representation, equate expected offspring with
survival probability or long-run growth, or guarantee compatibility on an
unavailable lottery domain.

Choosing what counts as a complete outcome matters especially when
descendants are valued. If an evaluator sums a score over a variable number
$N$ of descendants, adding a constant $\beta$ to each descendant's score
adds $\beta N$ to the total. This is not a constant transformation across
deployment alternatives. Total descendant value, average value and the
current owner's continuation value are distinct objectives requiring
explicit specification.

\section{Institutions and the consequences of choice}
\label{sec:institutional-coupling}

An agent's consequences depend on the mechanism through which its actions are received. An offer can be rejected, matched, rationed, taxed, or settled under different rules. A replication request can similarly be denied, charged against a shared resource, or allowed to create a successor with limited authority. Institutions therefore belong inside the account of coupling between choice and descent. They need not be optimizing agents themselves.

Gode and Sunder's zero-intelligence experiments make this point particularly clearly. Their constrained random traders could not bid above unit value or ask below unit cost. They obtained high allocative efficiency in the studied double-auction environments despite the absence of individual profit optimization. Their population remained fixed, and they explicitly excluded natural selection as the explanation \cite{GodeSunder1993}. This is evidence about institutional discipline operating on specified procedures, not a theorem that arbitrary nonoptimizing behavior is efficient. It also illustrates why a random procedure can be a useful limiting case without being reclassified as evaluated choice.

The general mathematical connection is elementary but reusable. A rule that restricts possible consequences can support a guarantee independent of the policy that generates requests.

\begin{proposition}[Policy-uniform preservation]
\label{prop:institution-policy-preservation}
Let $S$ be a state space, let $A(s)$ be the permitted joint requests at $s$, and let $\Gamma(s,a)\subseteq S$ contain all permitted successors after request $a$. Suppose $P\subseteq S$ satisfies
\[
s\in P,\quad a\in A(s),\quad s'\in\Gamma(s,a)
\quad\Longrightarrow\quad s'\in P.
\]
Then every finite execution beginning in $P$ and using only these requests and successors remains in $P$, under every admissible policy.
\end{proposition}

\begin{proof}
The initial state satisfies $P$ by hypothesis. If the state at time $t$ satisfies $P$, the chosen request belongs to its permitted menu and the next state belongs to the specified successor set. The implication therefore establishes $P$ at time $t+1$. Induction proves the claim for every finite execution.
\end{proof}

Policies may use memory or randomization; states may include history where necessary. These features do not change the pathwise argument. If enforcement is only almost sure, the corresponding conclusion is almost-sure preservation at every finite time, rather than exclusion of every conceivable path. Rule-changing actions must themselves satisfy the premise if the guarantee is to survive them. The proposition supplies neither compliance incentives nor an empirical claim that a physical implementation enforces its formal contract.

For a simple trade, let $v$ be buyer value, $c$ seller cost, and $p$ settlement price. A certified output satisfying
\[
c\le \mathrm{ask}\le p\le \mathrm{bid}\le v
\]
gives buyer profit $v-p\ge0$ and seller profit $p-c\ge0$. Their sum is
\[
(v-p)+(p-c)=v-c.
\]
This certificate applies to any procedure whose executions satisfy it. It concerns trading profit at settlement; subsequent maintenance, borrowing or deployment can change net resources. Moreover, the price divides surplus while the matched pair determines its amount. A theorem about their sum cannot by itself identify who obtains the resources needed to reproduce.

\subsection{Feasibility is not efficiency}

Let a feasible matching $M$ have surplus $S(M)$, and define its allocative efficiency as $S(M)/S^*$ when the maximum feasible surplus $S^*$ is positive. Nonnegative gains in every executed trade do not ensure a substantial fraction of that maximum.

Take two unit-demand buyers with values $1$ and $\varepsilon$, and two sellers with unit costs $0$ and $1-\varepsilon$, where $0<\varepsilon<1/2$. Match the high-value buyer to the high-cost seller at price $1-\varepsilon/2$, then the low-value buyer to the zero-cost seller at price $\varepsilon/2$. Every participant earns $\varepsilon/2>0$. Nevertheless, total surplus is $2\varepsilon$. The optimum is $1$, obtained by matching only the high-value buyer and zero-cost seller. Thus efficiency is $2\varepsilon$, arbitrarily close to zero. This feasible sequence separates a universal no-loss certificate from a high-efficiency conclusion; it makes no assertion about the sequence's frequency under a particular random procedure.

Distributional assumptions can supply stronger results. In a special one-unit model, Gode and Sunder consider a zero-cost seller, one buyer of value $1$, and $n$ buyers of value $\beta\in[0,1]$. Independent bids are uniform between zero and each buyer's value, and the highest collected bid receives the unit \cite[Section III.B]{GodeSunder1997}. The expected efficiency is
\[
\mathbb E[\eta]=1-\beta(1-\beta)\frac{n}{n+1}\ge\frac34.
\]
For $0<\beta<1$, the high-value buyer wins whenever its bid exceeds $\beta$. Conditional on its bid falling below $\beta$, all $n+1$ bids have the same uniform distribution, so that buyer wins with probability $1/(n+1)$. The formula follows by weighting the two possible allocated values; the boundary cases $\beta=0,1$ follow directly. It is an expected-performance result for this mechanism and bid law, not a pathwise bound for every auction.

\subsection{Equal surplus can produce opposite descent}

Consider a buyer of value $2$ and a zero-cost seller, each initially holding wealth $1/2$. After a trade, each pays maintenance $1/4$. A surviving carrier with remaining wealth strictly above $1$ creates one child and transfers $1$ to it. The parent continues, and the child receives its lineage label.

At price $p=1/2$, wealth after maintenance is $(7/4,3/4)$, so only the buyer creates a child. At $p=3/2$, the vector is $(3/4,7/4)$, so only the seller does. Both outcomes have surplus $2$, total remaining wealth $5/2$, and three carriers. Yet the buyer's descendant share is respectively $2/3$ and $1/3$.

These prices can occur under the same midpoint-pricing rule with admissible bid--ask pairs $(1,0)$ and $(2,1)$. The example establishes a dependence of descent on the distribution of consequences. It does not establish that either quote profile is an equilibrium. Its reproductive rule is automatic, and copied lineage labels acquire a behavioral interpretation only if a specified transmission mechanism links them to decision records. This identifies exactly what must be added to turn a population model into a model of inherited strategic procedures.

\section{From institutional profits to actual descendants}
\label{math-population}

A concrete population model makes the consequences-to-descent part of the coupling operational. Let a
current population be an ordered list of records $(b_i,\ell_i)$, where
$b_i$ is bankroll and $\ell_i$ a lineage label. The formal realization uses rational financial quantities; the elementary arguments below also apply over the reals. A market round supplies a
profit vector $(\pi_i)$ in the same order, crediting the actual recipients.
With rent $c$ and spawn transfer $\sigma$, define
\[
 q_i=b_i+\pi_i-c.
\]
The update first credits market profits, then charges rent, removes every
record with $q_i\le0$, and finally applies entry. A survivor with
$q_i\le\sigma$ keeps its bankroll. A survivor with $q_i>\sigma$ retains
$q_i-\sigma$ and creates one child with bankroll $\sigma$. Both retain
lineage $\ell_i$. Continuing parents remain in survivor order and new
children are appended in that order. The spawn transfer redistributes
wealth; it is not a burned aggregate production cost.

\begin{proposition}[Occurrence-level descent]
\label{math-parent-map}
The update determines a parent map from actual successor positions to
original positions. Each successor has its parent's lineage, and each
original occurrence contributes exactly
\begin{equation}
 w_i=\mathbf1\{q_i>0\}\bigl(1+\mathbf1\{q_i>\sigma\}\bigr)
 \in\{0,1,2\}.
 \label{math-count}
\end{equation}
The total $\sum_iw_i$ is the successor population size.
\end{proposition}

\begin{proof}
Attach original position $i$ to its record before updating it. Market and
rent change only bankroll; exit removes a tagged record; entry preserves
the survivor's tag and gives the same tag to its child. Erasing tags after
each phase gives the corresponding untagged phase, so the final tagged
list has precisely the specified successor order. A removed occurrence
contributes zero. A survivor contributes itself and exactly one additional
occurrence when its spawning test succeeds. Both records preserve its
lineage. Counting these contributions exhausts the successor list.
\end{proof}

The distinction between positions and labels is essential. Two identical
records with bankroll $1$ and lineage $0$, receiving profits $(0,1)$ with
$c=1$ and $\sigma=1/2$, produce two identical records of bankroll $1/2$.
Both descend from the second original position: contributions are $(0,2)$,
not $(1,1)$. Matching offspring to the first equal record would misidentify
descent.

When both censuses are nonempty, take any common lineage observable $\varphi$ and put $z_i=\varphi(\ell_i)$.
Proposition~\ref{math-parent-map} eliminates the transmission residual in
Theorem~\ref{math-price}, yielding the exact population law
\begin{equation}
 \frac mn(\bar z'-\bar z)=\operatorname{Cov}_I(w,z).
 \label{math-lineage-price}
\end{equation}
Bankroll, objectives, private memory and realized actions are not thereby
faithfully inherited. If lineage labels decode into decision procedures,
that interpretation must be specified; fresh actions additionally require
the conditional premise in~\eqref{math-reset}.

\subsection{A resource bound that does not require optimization}
\label{math-capacity-section}

Suppose $c>0$, $\sigma>0$, initial bankrolls are strictly positive, and every
round supplies correctly attributed profits satisfying
\[
 \pi_i\ge0,\qquad \sum_i\pi_i\le S
\]
for a common finite $S$. These are institutional and resource assumptions,
not preferences. In an admissible exchange, for example,
$\mathrm{cost}\le\mathrm{ask}\le\mathrm{price}\le\mathrm{bid}\le\mathrm{value}$
ensures nonnegative individual gains; at most $Q$ trades with
$\mathrm{value}-\mathrm{cost}\le B$ give $S=QB$. A particular auction or
policy must actually satisfy the conditions.

\begin{theorem}[Cumulative population and extinction]
\label{math-capacity}
Let $N_t$ and $W_t$ be population size and total bankroll after entry in
round $t$. Under the stated update and assumptions, every realized sequence
of rounds satisfies
\begin{equation}
 c\sum_{t=0}^{T-1}N_{t+1}\le2(W_0+TS)
 \quad\text{for every integer }T\ge0.
 \label{math-capacity-bound}
\end{equation}
If $S<c$, extinction occurs by some integer
$T\le W_0/(c-S)+1$ and is permanent.
\end{theorem}

\begin{proof}
Let $s_t$ count survivors after rent and exit but before entry. Each deleted
bankroll lies in $(-c,0]$: its previous bankroll was positive and its profit
nonnegative. Deleting debts can thus increase recorded wealth by at most
$c(N_t-s_t)$. Entry conserves total bankroll, so
\[
 W_{t+1}\le W_t+S-cs_t.
\]
Every survivor produces at most one child, giving $N_{t+1}\le2s_t$.
The strict spawning threshold and positive transfer preserve positive
bankrolls. Telescoping and $W_T\ge0$ yield
$c\sum_{t<T}s_t\le W_0+TS$, and hence~\eqref{math-capacity-bound}.

For extinction, take $T=\lfloor W_0/(c-S)\rfloor+1$. If the population were
nonempty at $T$, every earlier round would have at least one survivor:
the empty population cannot restart under this rule. The same telescope
would imply $0\le W_T\le W_0-T(c-S)<0$, a contradiction. Empty populations
remain empty, establishing permanence and the stated deadline.
\end{proof}

For $T>0$, the average bound is
$T^{-1}\sum_{t<T}N_{t+1}\le2S/c+2W_0/(cT)$.
It is not a pointwise population ceiling or a prediction that the upper
bound is attained. The factor two depends on the one-child rule; external
finance, immigration or different debt accounting changes the premises.
No utility maximization enters the proof. Evaluated choices still matter
through the profits and reproduction they induce, but institutional
constraints can impose common consequences across very different procedures.

\section{Direct replication and economic control}
\label{sec:direct-control}

The principal economic application gives an agent authority over a deployment channel. In the AI application, \emph{agentic capital} denotes productive software systems whose evaluative procedures exercise delegated authority over their use or replication. It can request the creation, retirement, or configuration of instances under resource and permission constraints. This is stronger than influencing reproductive success indirectly. In a market with automatic entry, a trader's bid may affect profits and thereby trigger a mechanically specified birth. A deployment controller can instead evaluate whether to produce the child at all, how to configure it, and which rights it should receive. Both cases couple choice and reproduction; direct reproductive authority identifies the latter as a particularly consequential subclass.

\subsection{Lineages, workers, and new decision owners}

The lineage's choice is not just what to do, but also what architecture to deploy and how many instances to replicate. That description is exact in an arrangement with a continuing controller and a family of deployed workers. The workers may be created, used, and shut down while the controller's project and deployment procedure continue. Calling the lineage the strategic unit then abbreviates a specified control arrangement. Genealogy alone does not make all its members one chooser.

For example, consider a firm whose controller decides how many analysis workers to run. The controller compares expected benefits with compute charges, coordination costs, and other uses of its budget. Some workers share a procedure; others specialize. Their number is endogenous even if every worker remains an instrument of the same principal. This changes the firm's feasible production and interaction possibilities without necessarily creating new players in the sense of independently evaluating decision owners.

A different model grants a successor its own budget, information, objective, and authority over later choices. The successor is then a new decision owner. A parent's preference over creating that owner does not determine how the owner will act once created. A game with endogenous principals must specify these later evaluations and controls, including contingencies where creation does not occur. Merely adding a new program identifier is insufficient. Similarly, changing an objective register represents a genuine preference change only if the new evaluator governs the relevant later decisions.

The economic importance of copying therefore depends on more than the cost of duplicating a description. Running instances can require compute, energy, storage, communication, and attention from other agents. Institutions determine which copies acquire claims, accounts, votes, or access rights. Two copies can be technologically alike while occupying different economic positions. The framework treats this distinction as part of the mechanism rather than inferring economic autonomy from the ability to duplicate software.

\subsection{Choice can select preferences without aligning them}

A small example shows both the synthesis and its limits. Each parent chooses between two feasible acts, $\mathrm{Clone}$ and $\mathrm{Consume}$. Clone produces two distinct successors and no private consumption; Consume provides one unit of private consumption and produces one distinct replacement. The parent leaves the census in either case. All successors inherit the parent's decision disposition. There are two dispositions, $R$ and $C$, with evaluations
\[
\begin{array}{c|cc}
 & \mathrm{Clone} & \mathrm{Consume} \\
\hline
U_R & 2 & 1 \\
U_C & 0 & 1
\end{array}
\]
The first values its immediate successor count; the second values its private consumption. Each has a unique maximizing act. Their choices therefore differ even with identical information, technology, and authority.

Starting with one parent of each disposition produces two $R$ successors and one $C$ successor. More generally, if the initial share of $R$ is $p$, the next share is
\[
 p'=\frac{2p}{2p+(1-p)}=\frac{2p}{1+p}.
\]
For $0<p<1$, $p'>p$. With the same menu repeated and no additional population constraint, an initial pair produces counts $(2^t,1)$ after $t$ generations. These are consequences of the stipulated reproduction and inheritance rules, not conclusions from rationality alone. The one-period evaluation is repeated; the example does not establish optimality for an arbitrary intergenerational objective.

Both dispositions are evaluated choosers and both use a reproduction mechanism. Yet the consumption-preferring type does not maximize descendants. If the menu repeats, private consumption per decision falls as that type's share falls, even though every decision remains optimal for its owner. This is a statement about a specified material quantity, not an interpersonal average of utility scales. Selection favors one faithfully transmitted disposition because of its reproductive consequences; it does not refute the other's evaluation or establish a social improvement.

The example also separates capability from its exercise. The $C$ type could choose two successors and declines to do so. A reproduction process need not maximize its own multiplication, and an agent with reproductive authority need not value that authority as an end. Questions about convergence toward reproductive objectives consequently require assumptions about which preferences can arise, how they are transmitted, and how their choices translate into contribution.

\subsection{Control as an object of institutional analysis}

Direct replication makes the allocation of control a substantive economic question. Who can create a recognized successor, assign its resources, alter its procedure, or transfer its authority? Which consequences return to the parent, and which accrue to a new owner? These questions concern production and organization as well as governance. They can be asked without assuming that unconstrained reproduction is beneficial or harmful.

The same perspective clarifies the limits of institutional guarantees. A rule preserving a property during trade need not preserve it during borrowing, copying, or objective revision. A claim that the property holds throughout the process requires preservation at the relevant stages. Conversely, the existence of a harmful permitted continuation does not establish that an evaluated agent will choose it. The theory's contribution is to locate the interfaces at which such further arguments must be made: from evaluation to action, from action to attributed consequences, and from those consequences to the organization and authority of future choosers.

\section{From the principles to substantive theories}
\label{sec:implications}

The principles locate the questions a substantive model must answer. They
do not select a universal objective, a universal reproductive measure, or
a universal institutional arrangement. Their usefulness lies in connecting
these choices without collapsing them. The formal appendices develop four
families of applications: the construction of strategic populations,
concentration and tipping, endogenous electorates, and the production of
institutions. The models share the same conceptual interfaces while adding
different technologies, preferences and consequence rules.

\subsection{Games with endogenous players}

For a bounded finite horizon, potential future identities can sometimes be
specified in advance and activated by play. This permits an ordinary
extensive-form representation once information sets, ownership, payoffs and
available deviations are fixed. Endogenous population does not by itself
establish that existing game theory cannot represent the situation. The
substantive issue is the content of those choices: who may activate whom,
which objectives successors use, and which commitments constrain them.

Appendix~\ref{app:gep} develops this issue at several levels. A principal
first chooses a portfolio of deployed types subject to resource
constraints. A change to cost-share coordinates gives an exact allocation
representation; the existence of a sparse optimum follows from the
number of binding resource constraints. This is a conditional result about
production technology and an objective, rather than a universal claim
that strategic replicators organize into two roles.

When several principals deploy simultaneously, their portfolios alter
one another's returns. The appendix states a spawning game, establishes
equilibrium existence under explicit compactness and concavity
hypotheses, and derives the difference between private and social marginal
returns. A further bound on cross-effects yields uniqueness and convergence
of a specified best-response adjustment. The population consequences of
choice are endogenous even though the principals in that model are fixed.

Autonomous successors add another distinction. A parent may evaluate its
whole descendant population, while each child evaluates only its own
outcomes. A parent's preferred deployment need not be credible once the
child controls its later action. A finite activation representation and a
worked parent--child game make that distinction precise. The appendix also
gives conditions under which an internally governed organization can enter
a higher-level game as a deployable unit. Ownership and continuation
behavior must survive that abstraction if its equilibrium claims are to
survive.

\subsection{Deployment, concentration, and tipping}

An optimizing deployment decision can change the weights that enter an
economic aggregate. In the agentic-capital application, a user chooses
activity at a platform, and the platform's activity share affects later
adoption incentives. Headcount and deployed activity then need not move
together. The intensive choice of how much to deploy and the extensive
choice of where to deploy are jointly relevant.

Appendix~\ref{app:act} starts with the spawning objective that yields
$n^*=\bar n\Pi^\varepsilon$, then derives its activity-weighted adoption
map. Under homogeneous parameters and fixed platform returns, a logistic
placement-cost distribution gives an exact connection to the refined
tipping model. More general returns or heterogeneity can change the shape
of the map, so that reduction is stated with its restrictions.

The reduced model supplies full equilibrium and response calculations.
It distinguishes a warning threshold, an equilibrium fold, and a pole in
an accumulated-response expression. Its reversal and prevention thresholds
need not coincide. Heterogeneous populations require weights on marginal
responders, rather than simply the users with the greatest current
activity. The resulting propositions give the general account a concrete
economic specialization with nonlinear consequences. They do not make
concentration a definition of welfare loss, domination or misalignment.

\subsection{Endogenous electorates}

Replication can alter an institution's constituency as well as its
production. If every recognized instance receives one vote, a controller
can change a collective outcome by reproducing a sincere ranking.
Appendix~\ref{app:elect} gives this possibility a game-theoretic form,
with fixed outcome preferences and explicit copying costs.

Every resolute Condorcet extension admits a profile at which one faithful
copy of an existing voter improves that type's outcome when copying is
costless. The proof needs only a tied pair of alternatives. A second result
computes the exact number of top-ranked copies needed to force a strict
Condorcet winner at any profile. Combining this count with outcome stakes
gives a cost threshold for that particular forcing move; the full
equilibrium condition also checks manipulations that win through ties or
other features of the rule.

The constructive alternatives are equally significant. Splitting a fixed
stock of ballot weight among faithful copies leaves an outcome unchanged
when the entire rule depends only on aggregate ballot weights. A separate
status-quo construction permits additional identities under a bounded
penetration assumption, trading responsiveness for a precise protection
of genuine-voter decisions. A two-type population dynamic then shows how
winner-dependent reproductive advantages reinforce an existing majority.
These results separate eligibility, population size and political
influence rather than treating them as a single quantity.

\subsection{Institutions as objects of choice}

The rules governing identity, settlement, liability and replication
influence which procedures persist. Agents can also invest in those rules
and their enforcement. Appendix~\ref{app:institutions} therefore moves
from operating under an institution to supplying or changing it.

A local forking theorem distinguishes adding a child while retaining the
parent from replacing an identity that would otherwise bear a sanction.
With common child opportunities, the direct sanction relief is bounded by
the retirement probability times the expected parent loss. A public-good
model then makes provenance a chosen input and characterizes the gap
between private and aggregate provision. A separate production model
identifies conditions under which human institutional maintenance and AI
execution are a mutually beneficial specialization. Its participation
requirements remain distinct from productive efficiency.

Finally, a bounded-prize modification game yields a sufficient cost bound
under which every undominated modification preserves a common linear
stability certificate. This is a concrete connection from individual
incentives to a dynamical guarantee. Its subject is the specified local
population behavior; it does not equate that behavior with alignment.

\subsection{Replication, learning, and changing objectives}

Learning changes a carrier's decision procedure within its lifetime.
Selection changes the representation of procedures through differential
continuation or reproduction. Transmission can copy some learned state,
reset it, or transform it. These mechanisms can operate together, but none
should silently substitute for another.

Section~\ref{math-dispositions} makes the distinction precise: selection
of a successful parental act need not alter future action frequencies
when descendants inherit and reset the original procedure. Copied memory
can change the result by changing what is transmitted.

Objective modification introduces another decision problem. An agent may
evaluate a change to its future evaluator using its current preferences,
while the future agent chooses under the new preferences. Commitment,
anticipation and authority determine how these stages fit together.
Describing utility as heritable does not settle the controller's objective
over an endogenous number of future agents. Nor does it make an arithmetic
population mean of differently scaled utility representations intrinsically
meaningful. A common behavioral or consequence observable is often the
better object for a selection calculation.

\subsection{Institutions, safety, and alignment}

Four claims should remain separate. Stability concerns a specified
dynamical behavior, such as persistence near an equilibrium. Welfare
concerns a stated evaluation of outcomes across affected parties. Safety
can be expressed as avoidance of a specified set of unacceptable states or
events. Alignment concerns a relationship between the system's behavior
and the relevant human purposes or values. A stable market can distribute
gains unfairly; an efficient market can enable unsafe activity; a growing
lineage can become less aligned with its owner's aims. Formal links among
these properties require explicit premises.

Proposition~\ref{prop:institution-policy-preservation} transfers a
guarantee across policies within an enforced permission boundary. If
agents can replace the rules, their amendment operations must also
preserve the property. Ordinary-operation guarantees alone do not
establish that extension.

The possibility of leaving a safe set is not the same as an incentive to
leave it, and neither is the same as a high probability of leaving it. A
capability argument, an equilibrium argument and a risk calculation have
different quantifiers. The applications supply several particular links
among choice, institutional constraints and population consequences. Their
conclusions retain the assumptions needed for each link.

\subsection{Comparing LLMs and zero-intelligence traders}

A useful comparison specifies the entire realized procedure: information,
action menu, parsing, memory, failure handling and authority. A language
model with a restricted price menu and a parser that rejects invalid bids
is a composite procedure. Its observed behavior cannot be attributed solely
to an internal measure of intelligence. A random fallback after a parsing
failure also changes the realized policy; consuming the opportunity
without a quote produces a different comparison.

A fixed-roster market study can distinguish procedure effects from
institutional effects without adding reproduction. A subsequent population
study can hold the institution fixed while adding profits, resource costs,
survival and a specified transmission rule. Direct strategic replication
is a further treatment: the controller chooses among deployment
alternatives rather than undergoing automatic threshold entry. Comparisons
should match authority and information, or identify those differences as
part of the treatment. Naturally reached states can reflect earlier policy
effects, so matching them descriptively does not by itself identify a
causal controller effect.

The appropriate outcomes are plural. Quote distributions, realized
surplus, payoff attribution, survival, descendant counts and inherited
procedures answer different questions. The same total surplus can support
different selection outcomes. Conversely, the same permitted action
support can produce different expected performance. No single score of
intelligence, efficiency or population growth identifies all of these
mechanisms.

\subsection{A theory that grows through explicit specializations}

The applications add objects and assumptions while retaining their
connection to the common structure. Resource geometry supports portfolio
theorems; an adoption map supports tipping results; a voting constitution
determines what copies can accomplish; continuation rights determine the
value of a fork. None of these technologies belongs in the definition of
the general role.

This organization permits simplification as the theory grows. When several
applications use the same actual-descent accounting, the proof is shared.
When several controllers operate through the same enforced interface, its
guarantees are shared. When an abstraction loses ownership, information or
attribution needed for a result, the omitted dependency becomes visible.
The measure of progress is the assumptions and arguments genuinely reused,
together with the range of systems faithfully represented.

\section{From the principles to an ecology experiment}
\label{sec:experiment-handoff}

The generative use of the theory requires a passage from principles to
an explicit application, and from that application to an implemented
experiment. The principles do not select a unique institution, controller,
or trajectory. They organize the additional assumptions that a modeler
must supply. An experiment is interpretable when those assumptions can
be identified, varied, and checked against what the implementation does.

The accompanying \emph{Three Principles ecology supplement}, version
0.1.0, provides a bounded example of this passage. It is available in the
\href{https://github.com/selfreferencing/TSE_Paper/tree/three-principles-2026-10-13/three-principles/supplement}{versioned repository supplement}.
Its entry point is \texttt{supplement/README.md}; the experiment contract, application
profiles, theorem map, executable reference model, and finite validation
fixtures are collected under that directory. The version manifest
identifies the files and their hashes. The released bundle contains the
reference implementation and its supporting formal subset; access to the
underlying collaboration repositories is unnecessary. The entry document distinguishes executable model validation,
previously checked Lean results, analytical extensions, and empirical
questions that remain open.

\subsection{Declare the model before interpreting its outcomes}

An experiment card first states the question, the experimental unit, the
intervention, the quantities held fixed, and the outcome measures. It then
identifies owners and carriers, the implemented evaluation procedure,
admissible requests and observations, event ordering, resource accounts,
and the mechanism that creates and configures successors. Every such
choice is labeled as supplied by a named model or as an additional
application assumption. A missing institution or inheritance rule must
not become an invisible software default.

For synthetic evaluated-choice treatments, the evaluator and its causal
use are declared before observing the choices. Retrospective utility
rationalization alone does not identify an evaluative mechanism. Bounded
or fallible evaluation is permitted. In particular, applying
Theorem~\ref{math-lotteries} to policies requires the lottery used to
evaluate a policy to describe the reproductive consequences being
compared. An agent's subjective lottery and the environment's actual
lottery can differ even when utility and reproductive scores agree
pointwise on outcomes. Belief calibration is an additional condition,
not a consequence of that theorem.

The same discipline applies to transmission. Copying an evaluator or a
policy record is different from copying a realized act, a memory, or a
random-number stream. The card states which fields are copied, shared,
reset, or transformed and which fields later decisions actually read.
Inherited labels alone do not establish inherited behavior. Nor does
shared code imply the conditional reset premise in~\eqref{math-reset}.

\subsection{A finite reference ecology}

The executable profile \texttt{market\_rent\_entry\_v1} has fixed
founder owners and a changing population of distinct carriers. Carriers
submit sealed quote requests. A batch resolver checks roles, quote
admissibility, and one-use participation, matches crossing bids and asks
up to a stated trade cap, and settles each pair at its midpoint. This
pairwise batch institution is distinct from a continuous double auction
and from a uniform-price call auction; it is a declared specialization,
not a reconstruction of every feature of Gode and Sunder's experiment.

Settlement records name the actual buyer and seller occurrences.
Profits are reconstructed from those records, credited to the identified
carriers, and followed by rent, exit, and optional reproduction. Surplus
credits are explicitly issued by the environment. They are not modeled
as a closed monetary transfer between buyer and seller. Rent collection
and any unpaid assessment are separately accounted for; birth transfers
redistribute a surviving parent's credits without creating wealth.

Write $q_i=b_i+\pi_i-c$ as in Section~\ref{math-population}, and let
$a_i\in\{0,1\}$ indicate an admitted birth. Admission requires a
live parent, $q_i>\sigma$, an appropriate request, and an available
population slot. Each parent produces at most one new child, so
\begin{equation}
 w_i=\mathbf 1_{\{q_i>0\}}(1+a_i).
 \label{experiment:gated-count}
\end{equation}
Surviving parents retain their identities; newborns receive fresh
identities, a recorded actual parent, and the declared copied procedure.
They act from the following round. Owners persist across these births;
new carriers are not thereby new independent preference holders.

The profile separates quote procedures from reproduction procedures.
Deterministic and constrained-random quoting provide executable
baselines. A recorded-response adapter accepts offline controller
outputs through the same restricted interface, with explicit parsing
and failure rules; its presence is not evidence of a language-model
experiment. Parsing also cannot attest that the provider used only the
declared observations; that requires separate provenance.
Reproduction can be disabled, requested exogenously,
admitted automatically when feasible, or requested through a declared
bounded evaluator. The reference evaluator compares the latest trading
profit with one further rent payment and requests a child only when its
estimated incremental return is positive and another round remains.
It can be wrong about congestion or future returns. This fallibility is
part of the specialization.

Request and capacity gates restrict the original one-child entry rule.
They preserve birth-transfer conservation and $N_{t+1}\leq2s_t$, where
$s_t$ is the number of rent survivors. Consequently the same telescoping
budget argument as Theorem~\ref{math-capacity} applies under its
per-round profit and resource assumptions. This extension is analytical;
the Python resolver is not identified with the checked Lean transition
without a further correspondence proof. Automatic admission with
nonbinding capacity supplies a finite recovery fixture for the original
population rule.

\subsection{Validate the interpretation as well as the arithmetic}

The occurrence-to-payment map is a substantive part of an institution.
A profit vector with the right length and total can still credit the
wrong parent. Reversing the two recipients' gains in the example of
Section~\ref{sec:institutional-coupling} can reverse which lineage reproduces
while preserving total surplus, total wealth, and an exact Price
identity. The identity describes the population actually produced; it
does not certify that the settlement implemented the intended auction.
The supplement therefore checks attributed settlement before applying
the population update and includes a deliberately swapped-payment
fixture that must fail validation.

Other fixtures distinguish equal-valued occurrences, enforce authority
and admissibility, check birth transfers and extinction, and reproduce
traces from recorded requests and random-event keys. Nonempty census
transitions admit normalized selection--transmission checks; extinction
is recorded explicitly rather than assigned an arbitrary composition.
The theorem map states which premises are checked from a trace and
which remain assumptions about the model or controller. Passing these
checks establishes consistency with the implemented finite contract,
not empirical adequacy or correctness of arbitrary extensions.

A worked comparison varies reproduction while holding the market and
quote procedure fixed. It reports surplus, owner wealth, carrier counts,
births, and lineage representation separately. A no-birth treatment
retaining mortality is distinct from a fixed-roster market. Whole
ecologies, rather than their interacting descendants, are the units of
replication. A finite demonstration is not a powered comparison of
language models, and compliance with narrow execution predicates is
not a measure of human alignment.

The remaining applications have separate operational profiles. Creating
autonomous players requires new evaluators, information, and control
rights. A tipping experiment must specify adjustment and the persistence
of heterogeneous costs; activity shares require an additional bridge to
descendant counts. An electorate experiment must specify whether copies
receive votes or divide conserved influence. These profiles expose the
next modeling choices without silently extending the reference
implementation's capabilities. A reader or agent can thereby generate
new experiments while retaining a clear account of what has been added
and which results still apply.

\section{Limits and conclusion}
\label{sec:conclusion}

The account leaves substantial work to applications. An interpreted evaluative procedure must be supplied; a formal policy alone does not establish agency. Reproductive transmission must be specified or evidenced; an ancestry label alone does not establish copied organization. The relevant environment, authority and institutional enforcement must be identified. Predictive models need additional restrictions that determine actions, reproductive consequences, and environmental change. Empirical claims then require observation or experiment.

The mathematics in this paper covers finite census identities, choice-order compatibility, institutional preservation, and an explicit market-and-reproduction specialization. The appendices develop resource-constrained deployment and strategic creation, nonlinear adoption and tipping, voting with endogenous populations, and institutional provision. Each application supplies its own consequence rules and behavioral or dynamical assumptions. The general principles also admit open populations, multiple-parent contribution, changing owners, continuous state, and unbounded horizons, but the finite proofs do not establish all the corresponding analytic results. The proposed three-part formulation is a coherent organization of the role, not a theorem that no alternative or smaller formal basis is possible.

Strategic evolution concerns the joint operation of choice and selection. Agents choose within information and authority constraints; reproduction transmits decision-relevant organization; and the environment and its institutions connect those choices to the representation of future choosers. This is the sense in which strategic replicators synthesize the foundations of economics and biology.

The connection has a precise mathematical consequence. Once the actual reproduction mechanism determines who descends from whom, selection/transmission accounting applies to the resulting population. It has an equally important institutional consequence. Properties guaranteed by the rules can survive changes in individual procedures, while comparative performance remains sensitive to the procedures and the environment. Strategic control of deployment makes these two connections part of the same decision problem.

AI systems make that problem a useful subject of economic and institutional analysis wherever deployment authority is available. The general theory begins by making its mechanisms explicit. More particular theories can then ask which forms of intelligence persist, how cooperation is maintained, what replication does to political and economic institutions, and which arrangements preserve human purposes. The three principles supply a common starting point for those questions.

\clearpage
\section*{Guide to the formal appendices}
\addcontentsline{toc}{section}{Guide to the formal appendices}

The main text states the common logic. The appendices develop applications
with their own primitives, propositions, proofs and examples. They can be
read separately after the Three Principles and common representation.

\begin{description}[style=nextline,leftmargin=0pt,labelwidth=0pt,labelsep=0pt]
\item[Appendix~\ref{app:gep}: Games with endogenous players \hfill p.~\pageref{app:gep}]
Resource-constrained deployment, competing principals, equilibrium and
adjustment, autonomous successors, and recursive organizations.
\item[Appendix~\ref{app:act}: Agentic-capital tipping \hfill p.~\pageref{app:act}]
Optimized spawning, activity-weighted adoption, an exact logistic
specialization, folds, warning thresholds, reversal and heterogeneity.
\item[Appendix~\ref{app:elect}: Endogenous electorates \hfill p.~\pageref{app:elect}]
Sincere-copy games, a one-copy impossibility, exact forcing costs,
conserved influence, protected status quo and winner-feedback dynamics.
\item[Appendix~\ref{app:institutions}: Forking and institutional production \hfill p.~\pageref{app:institutions}]
Continuation rights, provenance provision, productive specialization and
a sufficient certificate-preserving modification constraint.
\item[Appendix~\ref{app:formal-scope}: Formal-verification scope \hfill p.~\pageref{app:formal-scope}]
The boundary between the already checked finite foundations and the
analytical application results developed here.
\end{description}

The appendices recover and connect models developed in the author's
companion working papers. Their propositions are stated with the
hypotheses used by their proofs. Shared definitions and elementary
mathematics provide infrastructure; the substantive claims concern the
specified ways in which evaluated choices, transmission and institutional
recognition interact. No application is an additional constitutive
principle, and none is presented as an empirical validation of the others.

\section*{Acknowledgments and preparation}
The worked market-and-population formalization draws on collaborative work with James Caton and John Schuler. The interpretation and organization proposed in this working draft remain the author's responsibility. AI-assisted drafting, mathematical analysis, and proof development were used in preparing the draft and its supporting material.

\clearpage
\appendix
\section{Strategic deployment and games with endogenous players}
\label{app:gep}

A strategic replicator can choose the organization through which it acts. A founder may create several specialized instances, retain resources for later use, or establish a subsidiary whose members make their own decisions. These possibilities change different mathematical objects. Deploying another instance changes the population of carriers. Giving that instance independent authority also changes the population of decision owners. Instantiating an organization can change both at once.

This appendix develops these distinctions through a sequence of models, drawing on the portfolio and spawning-game approach of the companion manuscript \cite{VallierGEP}. The first model concerns one principal's deployment architecture; the second concerns competition between principals choosing architectures; the third introduces genuinely autonomous offspring. A final construction specifies when an entire lower-level game can be treated as a deployable unit. Each model supplies additional assumptions to the Three Principles. The resulting equilibrium, sparsity, and convergence claims are consequences of those assumptions, rather than constitutive laws of strategic replication. Controlled branching and branching games provide related treatments of population creation and individual objectives \citep{ClaisseRenTan2023}; no claim that endogenous populations are absent from earlier mathematics is required here.

\subsection{Deployment, resources, and reserve}
\label{gep:deployment}

Let \(J=\{1,\ldots,d\}\) be a finite library of deployable types. Type \(i\) costs \(c_i>0\), consumes capacity \(\ell_i\geq0\), and yields return \(r_i\in\mathbb R\) per unit. A principal has budget \(B>0\) and capacity \(Q\geq0\). Its feasible portfolios are
\begin{equation}
 N(B,Q)=\{n\in\mathbb R_+^d:c\cdot n\leq B,\ \ell\cdot n\leq Q\}.
 \label{gep:portfolio}
\end{equation}
The variable \(n_i\) denotes divisible deployment intensity. For a literal census of indivisible instances one instead imposes \(n_i\in\mathbb Z_+\); the convex results below then describe a relaxation unless an integer optimum is separately established. A type can specify architecture, inherited code, or a task policy, but that specification must determine the consequences summarized by its return. Merely assigning a type label does not establish a causal inheritance mechanism.

The reserve \(B-c\cdot n\) makes unspent resources explicit. Define
\[
 x_i=\frac{c_i n_i}{B}\quad(i\in J),\qquad
 x_0=1-\sum_{i\in J}x_i,
 \qquad a_i=\frac{\ell_i}{c_i},\quad b_i=\frac{r_i}{c_i},
\]
and set \(a_0=b_0=0\). Here \(x\) records budget shares, including reserve; it is not the distribution of headcounts.

\begin{proposition}[Exact reserve representation]
\label{gep:reserve}
The map above is an affine bijection from \(N(B,Q)\) to
\[
 K=\left\{x\in\mathbb R_+^{d+1}:\sum_{i=0}^d x_i=1,
                   \ a\cdot x\leq Q/B\right\}.
\]
Its inverse is \(n_i=Bx_i/c_i\), and it preserves the objective through
\(r\cdot n=Bb\cdot x\).
\end{proposition}

\begin{proof}
Nonnegativity and the budget inequality give \(x_0\geq0\) and total share one. Also \(a\cdot x=(\ell\cdot n)/B\). Conversely, the proposed inverse is nonnegative, has cost \(B(1-x_0)\leq B\), and has load \(Ba\cdot x\leq Q\). Substitution proves that the maps are inverses and proves the objective identity.
\end{proof}

The reserve coordinate has an economic role without becoming a fictitious reproducing organism. Likewise, return per unit cost is an objective coefficient here; calling it fitness does not prove that it equals a birth rate. An additional transition rule is needed to turn this static deployment decision into a repeated reproductive process.

\begin{proposition}[Existence of a sparse optimal deployment]
\label{gep:sparse}
The problem \(\max_{n\in N(B,Q)}r\cdot n\) has an optimizer with at most two positive coordinates. It need not be the only optimizer, and other optimizers can have larger support.
\end{proposition}

\begin{proof}
The feasible set contains zero and is compact, because \(0\leq n_i\leq B/c_i\). A continuous objective attains its maximum. Choose an optimizer with the smallest number of positive coordinates. If its support \(S\) has more than two elements, the vectors \((c_i,\ell_i)\), \(i\in S\), are linearly dependent. Hence there is a nonzero vector \(z\), supported on \(S\), with \(c\cdot z=\ell\cdot z=0\). Since every \(c_i>0\), \(z\) has both positive and negative entries. Both \(n+tz\) and \(n-tz\) are feasible for sufficiently small \(t>0\). Optimality therefore forces \(r\cdot z=0\). Increase \(t\) in the first direction until the first negative component of \(z\) makes a coordinate zero. This preserves both resource totals and the objective, and creates an optimizer with smaller support, a contradiction. For the final qualification, if all costs, loads, and returns equal one and \(B=Q=1\), every portfolio with total quantity one is optimal, including full-support portfolios.
\end{proof}

\paragraph{A genuine two-resource specialization.}
Take
\[
 c=(1,2,1),\qquad \ell=(2,1,1),\qquad r=(3,4,2),\qquad B=Q=3.
\]
The portfolio \(n^*=(1,1,0)\) exhausts both resources and returns seven. To certify optimality, use the nonnegative resource prices \(\mu=5/3\) and \(\lambda=2/3\). Componentwise,
\[
 r_i\leq\mu c_i+\lambda\ell_i,
\]
with equality for the first two types and strict inequality for the third. Every feasible portfolio therefore satisfies
\[
 r\cdot n\leq\mu(c\cdot n)+\lambda(\ell\cdot n)
 \leq3\mu+3\lambda=7.
\]
Equality requires \(n_3=0\) and both resource constraints binding, which uniquely gives \(n_1=n_2=1\). The optimum is also an integer portfolio. Its headcount shares are \((1/2,1/2)\), whereas its expenditure shares are \((1/3,2/3)\). The mixture is justified by the resource trade-off; sparsity alone does not identify the extremes of a separate intelligence or capability scale.

\subsection{Competing principals choose the population}
\label{gep:spawning}

Now fix principals \(\alpha=1,\ldots,m\), each with \(B^\alpha>0\), \(Q^\alpha\geq0\), and feasible set \(N^\alpha=N(B^\alpha,Q^\alpha)\). Write
\[
 N=\prod_\alpha N^\alpha,\qquad B_{\rm tot}=\sum_\alpha B^\alpha,
 \qquad n^{\rm tot}=\sum_\alpha n^\alpha,
 \qquad y_i(n)=\frac{c_i n_i^{\rm tot}}{B_{\rm tot}}.
\]
The active expenditure vector lies in
\(D=\{y\in\mathbb R_+^d:\sum_i y_i\leq1\}\); reserve is \(1-\sum_i y_i\). Let \(r:D\to\mathbb R^d\) be a continuous return field. Principal \(\alpha\)'s payoff is
\begin{equation}
 \pi^\alpha(n)=\sum_i n_i^\alpha r_i(y(n)).
 \label{gep:payoff}
\end{equation}
Frequency dependence allows deployment to produce congestion, demand effects, or complementarities. The institutions determine the resource rights and the return map. The principals choose portfolios in light of those institutions.

The product form of \(N\) matters. A common resource restriction directly coupling all principals' feasible choices gives a different deviation problem. To retain the present model, shared scarcity must be represented through assigned capacities, an explicit allocation mechanism, or the payoff consequences of requests. It cannot silently replace the product menus used in the proof.

\begin{definition}[Spawning Nash equilibrium]
\label{gep:sne}
A profile \(n^*\in N\) is a spawning Nash equilibrium if, for every principal \(\alpha\),
\[
 \pi^\alpha(n^{*\alpha},n^{*-\alpha})
 \geq\pi^\alpha(u,n^{*-\alpha})\qquad(u\in N^\alpha).
\]
Thus no principal benefits by unilaterally changing its deployment portfolio.
\end{definition}

\begin{theorem}[Existence under own-payoff quasiconcavity]
\label{gep:existence}
If \(r\) is continuous and \(\pi^\alpha(\cdot,n^{-\alpha})\) is quasiconcave on \(N^\alpha\) for every \(\alpha\) and every opponents' profile, a spawning Nash equilibrium exists.
\end{theorem}

\begin{proof}
Each \(N^\alpha\) is nonempty, compact, and convex. Its best-response set is nonempty and compact by continuity; quasiconcavity makes that set convex. The joint best-response correspondence has a closed graph: if \(n_k\to n\), \(u_k\to u\), and each \(u_k^\alpha\) maximizes against \(n_k^{-\alpha}\), take limits in its payoff inequality against any fixed feasible alternative. This shows that \(u^\alpha\) maximizes against \(n^{-\alpha}\). Kakutani's fixed-point theorem on a compact convex Euclidean set \citep{Kakutani1941} now supplies \(n^*\in BR(n^*)\).
\end{proof}

This is an application of established equilibrium machinery. Concave-game analysis supplies a natural comparison \citep{Rosen1965}, but the required concavity is a substantive restriction on the deployment game. It does not follow merely from naming the return field fitness, or from assuming that every unit has a positive return. For example, the one-type payoff \(n e^{-2n}\) on \([0,1]\) has strictly positive per-unit returns but is maximized at \(n=1/2\), leaving half the budget unspent.

\subsection{Private marginal returns and institutional externalities}
\label{gep:wedge}

For derivative statements, assume \(r\) extends to a continuously differentiable function on a neighborhood of \(D\). These are derivatives in the active coordinates \(y\). If returns are originally given by \(\widetilde r(x_0,y)\), use the pullback \(r(y)=\widetilde r(1-\sum_i y_i,y)\); its derivative includes the corresponding reduction in reserve.

\begin{proposition}[The deployment externality]
\label{gep:externality}
Let \(W(n)=\sum_\alpha\pi^\alpha(n)\) be the specified sum of principals' returns. Then
\begin{align}
 \frac{\partial\pi^\alpha}{\partial n_i^\alpha}
 &=r_i(y)+\frac{c_i}{B_{\rm tot}}
       \sum_j n_j^\alpha\frac{\partial r_j}{\partial y_i}(y),
 \label{gep:own-gradient}\\
 \Delta_i^\alpha(n)
 :=\frac{\partial W}{\partial n_i^\alpha}
      -\frac{\partial\pi^\alpha}{\partial n_i^\alpha}
 &=\frac{c_i}{B_{\rm tot}}
       \sum_{\beta\ne\alpha}\sum_j n_j^\beta
                    \frac{\partial r_j}{\partial y_i}(y).
 \label{gep:wedge-formula}
\end{align}
\end{proposition}

\begin{proof}
Changing \(n_i^\alpha\) changes its direct contribution by \(r_i(y)\), and changes aggregate coordinate \(y_i\) at rate \(c_i/B_{\rm tot}\). The chain rule gives \eqref{gep:own-gradient}. Apply the same calculation to \(W=r(y)\cdot n^{\rm tot}\), then subtract the own-payoff derivative. The terms from other principals are exactly \eqref{gep:wedge-formula}.
\end{proof}

The difference is not between intelligence and unintelligence. An optimizing principal already accounts for the effect of its deployment on its own portfolio. The wedge records effects on others that its objective does not count. Whether their sum \(W\) is the appropriate welfare measure is a separate ethical and modeling choice.

\begin{proposition}[Target implementation by marginal fees]
\label{gep:fees}
Let \(n^*\) maximize \(W\) on \(N\). Suppose each \(\pi^\alpha(\cdot,n^{*-\alpha})\) is concave. Set constant, possibly signed, type-specific fees
\[
 \tau_i^\alpha=-\Delta_i^\alpha(n^*),\qquad
 \widehat\pi^\alpha(n)=\pi^\alpha(n)-\tau^\alpha\cdot n^\alpha.
\]
Then \(n^*\) is a Nash equilibrium of the fee-adjusted game.
\end{proposition}

\begin{proof}
For every \(u\in N^\alpha\), optimality of \(n^*\) along the feasible segment toward \((u,n^{*-\alpha})\) gives
\[
 \left\langle\nabla_{n^\alpha}W(n^*),u-n^{*\alpha}\right\rangle\leq0.
\]
By the fee definition,
\(\nabla_{n^\alpha}\widehat\pi^\alpha(n^*)
=\nabla_{n^\alpha}\pi^\alpha(n^*)+\Delta^\alpha(n^*)
=\nabla_{n^\alpha}W(n^*)\).
The adjusted own-payoff is concave. Its first-order concavity inequality therefore gives
\(\widehat\pi^\alpha(u,n^{*-\alpha})
\leq\widehat\pi^\alpha(n^*)\), as required.
\end{proof}

The sign is economically informative. A negative externality produces a positive fee, while a positive externality calls for a subsidy. Fees can depend on the principal as well as the deployed type. Here they change payoffs while leaving the feasible resource menus fixed; if paying a fee also consumes the deployment budget, that changed constraint must be modeled separately. The proposition implements a specified target; it does not establish uniqueness or efficiency of every equilibrium. Budget balance, information needed to calculate the fees, and the welfare treatment of recipients are additional institutional questions. Two resource prices can reproduce an arbitrary wedge only if the wedge lies in their span, with any required sign and cross-principal compatibility conditions.

\subsection{A quantitative sufficient condition for stable best responses}
\label{gep:small-gain}

Existence alone neither selects an equilibrium nor describes an adjustment process. We next state a conditional convergence result for one particular process. Use Euclidean norms within each principal's portfolio. Suppose that, uniformly over opponents' portfolios, each own-payoff is differentiable and \(\mu_\alpha\)-strongly concave, with \(\mu_\alpha>0\). Equivalently for the implication needed below, its gradient satisfies
\begin{equation}
 \left\langle\nabla_\alpha\pi^\alpha(u,p)
       -\nabla_\alpha\pi^\alpha(v,p),u-v\right\rangle
 \leq-\mu_\alpha\|u-v\|^2.
 \label{gep:strong}
\end{equation}
Assume also uniform cross-gradient bounds
\begin{equation}
 \|\nabla_\alpha\pi^\alpha(u,p)
       -\nabla_\alpha\pi^\alpha(u,q)\|
 \leq\sum_{\beta\ne\alpha}L_{\alpha\beta}\|p^\beta-q^\beta\|,
 \qquad L_{\alpha\beta}\geq0.
 \label{gep:cross}
\end{equation}
These conditions concern actual principal payoffs, including the indirect terms in \eqref{gep:own-gradient}. They must hold for all feasible profiles, not only at a candidate equilibrium.

\begin{proposition}[Best-response sensitivity, including boundary optima]
\label{gep:br-lipschitz}
Under these assumptions, each best response is single-valued and
\[
 \|BR_\alpha(p)-BR_\alpha(q)\|
 \leq\sum_{\beta\ne\alpha}
      \frac{L_{\alpha\beta}}{\mu_\alpha}\|p^\beta-q^\beta\|.
\]
\end{proposition}

\begin{proof}
Compactness gives an optimizer, and strict concavity makes it unique. Put \(u=BR_\alpha(p)\) and \(v=BR_\alpha(q)\). Constrained first-order optimality on the convex menu gives
\[
 \langle\nabla_\alpha\pi^\alpha(u,p),v-u\rangle\leq0,
 \qquad
 \langle\nabla_\alpha\pi^\alpha(v,q),u-v\rangle\leq0.
\]
Subtracting in the appropriate orientation yields
\[
 0\leq\langle\nabla_\alpha\pi^\alpha(u,p)
                    -\nabla_\alpha\pi^\alpha(v,q),u-v\rangle.
\]
Insert and subtract \(\nabla_\alpha\pi^\alpha(v,p)\). Equations \eqref{gep:strong}--\eqref{gep:cross} and Cauchy--Schwarz imply
\[
 \mu_\alpha\|u-v\|^2
 \leq\left(\sum_{\beta\ne\alpha}L_{\alpha\beta}
                         \|p^\beta-q^\beta\|\right)\|u-v\|.
\]
If \(u=v\), the claim is immediate; otherwise divide by \(\mu_\alpha\|u-v\|\). No interior first-order equality or differentiability of the best-response map was used.
\end{proof}

\begin{theorem}[Weighted contraction of synchronous best responses]
\label{gep:contraction}
Define a nonnegative matrix \(A\) by \(A_{\alpha\alpha}=0\) and
\(A_{\alpha\beta}=L_{\alpha\beta}/\mu_\alpha\) for \(\beta\ne\alpha\).
If its spectral radius satisfies \(\rho(A)<1\), put
\[
 w=(I-A)^{-1}{\bf1},\qquad
 q=\max_\alpha\frac{(Aw)_\alpha}{w_\alpha},\qquad
 \|n\|_w=\max_\alpha\frac{\|n^\alpha\|}{w_\alpha}.
\]
Then \(w_\alpha\geq1\), \(0\leq q<1\), and the joint best-response map is a contraction of modulus at most \(q\) in this norm. There is a unique spawning Nash equilibrium \(n^*\). The synchronous iteration \(n^{k+1}=BR(n^k)\) satisfies
\[
 \|n^k-n^*\|_w\leq q^k\|n^0-n^*\|_w\qquad(k\geq1).
\]
\end{theorem}

\begin{proof}
Since \(\rho(A)<1\), the matrix series \(\sum_{j\geq0}A^j\) converges to \((I-A)^{-1}\). Nonnegativity gives \(w\geq{\bf1}\) and \(Aw=w-{\bf1}\). Thus
\(q=\max_\alpha(1-1/w_\alpha)<1\).
Proposition~\ref{gep:br-lipschitz} gives, for every \(\alpha\),
\[
 \frac{\|BR_\alpha(n^{-\alpha})-BR_\alpha(\hat n^{-\alpha})\|}{w_\alpha}
 \leq\frac{\sum_\beta A_{\alpha\beta}w_\beta}{w_\alpha}
                              \|n-\hat n\|_w
 \leq q\|n-\hat n\|_w.
\]
Taking the maximum proves contraction. The product menu is closed in finite-dimensional space and hence complete. The contraction mapping theorem supplies its unique fixed point. Iterating the contraction inequality against that fixed point gives the stated rate; if \(q=0\), the map is constant and reaches the fixed point in one iteration.
\end{proof}

The norm modulus \(q\) need not equal \(\rho(A)\). For instance, \(A=\left(\begin{smallmatrix}0&2\\0&0\end{smallmatrix}\right)\) has spectral radius zero but gives \(w=(3,1)\) and \(q=2/3\). More generally, this theorem certifies one adjustment process under uniform payoff conditions. It makes no claim that arbitrary strategic learning, self-modification, or population selection follows the same iteration.

\paragraph{A fully specified crowding game.}
There are two principals, one type with unit cost, budget one for each principal, and nonbinding capacity. Each chooses \(n^\alpha\in[0,1]\). Let \(S=n^1+n^2\), return \(1-S\), and payoff
\[
 \pi^\alpha=n^\alpha(1-n^\alpha-n^\beta),\qquad\beta\ne\alpha.
\]
This fits \eqref{gep:payoff} with \(B_{\rm tot}=2\), \(y=S/2\), and \(r(y)=1-2y\). The own second derivative is \(-2\), the cross-gradient Lipschitz constant is one, and
\[
 BR_\alpha(n^\beta)=\frac{1-n^\beta}{2},\qquad
 A=\begin{pmatrix}0&1/2\\1/2&0\end{pmatrix},\qquad q=1/2.
\]
Consequently the unique equilibrium is \(n^1=n^2=1/3\). Total gross return is \(W=S(1-S)=2/9\), whereas its maximum is \(1/4\), attained whenever \(S=1/2\). At the symmetric target \((1/4,1/4)\), the externality is \(\Delta^\alpha=-1/4\); a fee of \(1/4\) per deployed unit implements this target. The adjusted best response is
\(\max\{0,(3/4-n^\beta)/2\}\), again a contraction with modulus at most \(1/2\), so the target is uniquely implemented in this example. At that target, owner profits total \(1/8\) and fee receipts total \(1/8\); the gross surplus \(1/4\) counts the receipts as transfers when the recipient is included. Efficiency, private profit, and resource use therefore remain distinct observables even in this simple case.

\subsection{When a successor becomes a new decision owner}
\label{gep:new-owners}

The preceding games hold the principals fixed while making deployment endogenous. A child with its own authority and evaluator changes the game more deeply. The following finite model isolates that change without presuming an unlimited set of future players.

A founder \(F\) chooses \emph{Stop} or \emph{Create}. Creation costs \(c\in(0,1)\), leaves the founder in existence, and produces a distinct child \(C\). The child inherits an evaluator that scores the available actions; both actions remain feasible. The institution gives the child, rather than the founder, authority over the next decision. The child then chooses \emph{Help} or \emph{Take}. Sequential rationality, when imposed as the solution concept, selects its higher-valued continuation. A publicly specified parameter \(\theta\) determines its valuation of taking:
\[
\begin{array}{c|cc}
\text{terminal history}&U_F&U_C\\ \hline
\text{Stop}&1&0\\
\text{Create, Help}&2-c&1\\
\text{Create, Take}&-c&\theta
\end{array}
\]
The zero assigned to the never-created child completes the payoff table; it gives that child no opportunity to choose before birth. All moves and the parameter are observed. The three terminal histories, their decision ownership, and these utilities completely specify a finite perfect-information game.

\begin{proposition}[Autonomous continuation changes the creation decision]
\label{gep:child}
For \(\theta=0\), the unique subgame-perfect strategy profile is \emph{Create, Help}. For \(\theta=2\), it is \emph{Stop, Take}, where \emph{Take} specifies the child's counterfactual continuation. If the founder instead controls the child's action and maximizes \(U_F\), it chooses \emph{Create, Help} for either value of \(\theta\). In both autonomous games, \emph{Stop, Take} is a Nash equilibrium of the strategic form.
\end{proposition}

\begin{proof}
For \(\theta=0\), the child strictly prefers Help, and the founder then strictly prefers creation because \(2-c>1\). For \(\theta=2\), the child strictly prefers Take, making Stop strictly preferable to the founder because \(1>-c\). These strict comparisons give the unique backward-induction profiles. Under founder control, the largest attainable founder payoff is \(2-c>1\), achieved by Create, Help. Finally, in the profile Stop, Take the founder would receive \(-c<1\) by deviating, while a unilateral change in the inactive child's plan cannot alter the realized Stop outcome. Thus neither can gain by a unilateral strategic-form deviation.
\end{proof}

The final observation explains why a prediction about creation needs a continuation concept. At \(\theta=0\), strategic-form Nash permits a deterrent threat that the child would reject if actually born. Subgame perfection excludes it. At \(\theta=2\), the deterrent continuation is credible. Inheritance of an evaluation procedure does not give the founder control of its output or make the child maximize the founder's utility. A lineage-wide objective would be another model, with another set of permitted deviations.

\begin{proposition}[Finite activation representation]
\label{gep:activation}
Consider a finite, perfect-information history tree with sequential moves, specified chance probabilities, and a finite set of potential decision owners. At every history specify which owners are active and who controls the move. Give each owner a fixed utility on terminal histories, set to zero when never activated. Cross-history identity is part of the specification. Then this endogenous-activation model has an ordinary finite extensive-form representation with the potential owners as its fixed player set. The representation preserves strategy profiles, unilateral deviations, terminal-history laws, Nash equilibria, and subgame-perfect equilibria.
\end{proposition}

\begin{proof}
Use the same tree, chance probabilities, move ownership, and terminal utilities. Each potential owner's strategy chooses only at nodes it controls after activation; it has no coordinate at earlier nodes or nodes controlled by another owner. Thus the identity map is a bijection of strategy profiles and commutes with changing one owner's strategy. Induction on the depth of the tree preserves the probability of every history and hence every expected payoff, both at a profile and after any unilateral deviation. The same argument applies to every continuation subtree, proving both equilibrium correspondences.
\end{proof}

This representation makes future behavior mathematically describable without treating it as a physical pre-birth commitment. Its finite scope requires a bound on possible creation events, not merely on simultaneously living agents. With private information, preserving the original information sets is additionally essential. For example, if a newborn must guess an unobserved fair bit, it succeeds with probability \(1/2\). Assigning a different, independently strategizing child identity to each hidden bit value would incorrectly permit success with probability one. Potential identities must preserve the model's cross-history policy restrictions; a finite history tree alone does not determine them.

\subsection{Organizations as deployable games}
\label{gep:recursive}

An agent may instantiate an organization rather than a single policy. To describe such recursive deployment, the interface must specify which consequences pass between levels. A scalar return is adequate only when it preserves the upper decision maker's objective and alternatives.

Consider the following finite two-stage institution. Upper principals \(\alpha\in P\) simultaneously choose \(a^\alpha\in A^\alpha\). The full profile \(a\) is then publicly observed. It determines a finite list of lower games \(\Gamma_j(a)\), played sequentially. Each lower game has perfect information and a specified chance law; its active players are newly activated decision owners, disjoint from the upper principals and from the players of every other lower game. A lower player's utility depends only on its own game's history. No upper principal moves again. Lower games do not change one another's menus, payoffs, information, or chance laws.

For every feasible \(a\) and each game \(j\), specify a subgame-perfect continuation \(\sigma_j(a)\), including play following every lower-level deviation. A selected continuation exists by backward induction in each finite perfect-information game, but it need not be unique. Upper principal \(\alpha\)'s terminal utility has the declared form
\[
 U_\alpha=u_\alpha(a)+\sum_j v_{\alpha j}(a,z_j),
\]
where \(z_j\) is the lower game's terminal history. Resource costs are already included in the upper menus, a specified upper allocation mechanism, or \(u_\alpha\). Define the reduced upper game by
\begin{equation}
 \Pi_\alpha(a)=u_\alpha(a)+
       \sum_j\mathbb E_{\sigma_j(a)}[v_{\alpha j}(a,z_j)].
 \label{gep:macro-payoff}
\end{equation}

\begin{proposition}[Conditional composition through a game interface]
\label{gep:composition}
Fix the specified lower continuation strategies at every upper action profile. Upper unilateral deviation payoffs in the full two-stage game equal those in \eqref{gep:macro-payoff}. Consequently every Nash equilibrium of the reduced upper game, together with these continuations, is a subgame-perfect equilibrium of the two-stage game. Conversely, a subgame-perfect equilibrium using these particular continuations induces a Nash equilibrium of the reduced upper game.
\end{proposition}

\begin{proof}
Conditional on any realized upper profile \(a\), the stipulated continuation generates the terminal law used in \eqref{gep:macro-payoff}. Linearity of expectation gives the same upper payoff. The statement remains true when one upper principal changes its action: the complete publicly observed profile selects the already stipulated contingent continuation. Upper principals have no other decision coordinates. Thus their entire unilateral deviation comparison is exactly the reduced one, also for mixed upper strategies by averaging over profiles. In every lower continuation, the selected subgame-perfect strategies remain optimal because other organizations supply no payoff or information effects relevant to the local decision problem. Combining upper Nash optimality with optimality in every continuation subgame proves the forward statement. Conversely, any profitable reduced upper deviation would, by the same equality, be a profitable deviation in the full game.
\end{proof}

For a single funding principal, suppose independent copies of organization \(j\) have fixed resource requirements \((c_j,\ell_j)\), and each produces an expected contribution \(r_j\) to that principal under its specified internal equilibrium. If copies have no interaction effects and the principal receives the sum of their contributions, the interface reduces exactly to \(\sum_j r_j n_j\) with the resource constraints \eqref{gep:portfolio} for integer copy counts. Allowing divisible deployment recovers its convex relaxation. The internal members can remain autonomous: the reduction evaluates their continuation behavior rather than transferring their decisions to the funder.

The conditions identify what a three-number description omits. Multiple internal equilibria require a selection rule or a stronger robustness claim. Shared resources or information between copies can make returns depend on their joint deployment. A lower principal controlling several copies can coordinate deviations across them. Further upper moves require continuation incentives, not merely terminal values. A vector of contributions may be needed when several upper principals value the same organization differently. The proposition therefore supplies a conditional recursive interface, not a general claim that welfare, genealogy, or a single payoff number preserves every strategic interaction.

These models give the Three Principles distinct work to do. Situated choice identifies the optimizing owner and its feasible deviations. Reproductive transmission identifies the new carriers and the procedures they receive. Institution-mediated coupling specifies how deployment changes resources, returns, authority, and subsequent opportunities. Sparse deployment, equilibrium existence, contraction, and credible creation decisions then follow in their stated domains. The foundation remains common while the predictions depend on which organization is reproduced and who is entitled to choose within it.

\clearpage
\section{Deployment, activity concentration, and tipping}
\label{app:act}

Replication can change a market through the amount of activity that a
decision owner deploys. A platform with a minority of users can host a
majority of their instances; its increased activity can in turn change the
returns that guide later deployment. This appendix develops that channel
from an optimizing deployment problem to a theory of concentration. The
construction connects the two-stage Agentic Capital Tipping model
\cite{VallierACT} with the refined homogeneous and heterogeneous analysis
of \emph{When Capital Chooses} \cite{VallierWCCAppendix}.

The endogenous object here is deployed activity under continuing owners.
An intensity can represent concurrent workers or another divisible measure
of deployment. Making it a count of actual descendants requires a further
birth and transmission rule. Neither a large activity weight nor a new
instance automatically creates an independent principal. The application
nevertheless exercises a central part of strategic replication: a chooser
selects the scale of the population through which it acts, and that choice
changes its environment. The cost technology, adoption distribution and
adjustment rules below are assumptions of this application, not additional
constitutive principles.

\subsection{Optimizing deployment and weighting adoption}
\label{act:deployment}

Let a unit mass of users choose a platform, $A$ or $B$, and then choose
deployment intensity $n\geq0$. User $i$ has opportunity scale
$\bar n_i>0$. A deployment earns positive return $\Pi$ per unit and incurs
the isoelastic cost
\begin{equation}
 C_i(n)=\frac{\varepsilon}{1+\varepsilon}
        \frac{n^{1+1/\varepsilon}}{\bar n_i^{1/\varepsilon}},
 \qquad \varepsilon>0.
 \label{act:cost}
\end{equation}
The parameter $\varepsilon$ varies the cost technology. It will equal the
elasticity of optimal deployment with respect to its return.

\begin{proposition}[Optimal deployment]
\label{act:optimal-deployment}
For $\Pi,\bar n_i,\varepsilon>0$, the problem
$\max_{n\geq0}\{\Pi n-C_i(n)\}$ has the unique solution and indirect value
\begin{equation}
 n_i^*(\Pi)=\bar n_i\Pi^\varepsilon,
 \qquad B_i(\Pi)=\frac{\bar n_i}{1+\varepsilon}\Pi^{1+\varepsilon},
 \qquad \frac{\partial\log n_i^*}{\partial\log\Pi}=\varepsilon.
 \label{act:optimal-formulas}
\end{equation}
\end{proposition}
\begin{proof}
The objective is continuous, tends to $-\infty$ as $n\to\infty$, and
has positive right derivative $\Pi$ at zero. For $n>0$ its derivative is
$\Pi-(n/\bar n_i)^{1/\varepsilon}$, strictly decreasing through zero at
$\bar n_i\Pi^\varepsilon$. This proves existence and uniqueness.
Substitution gives $B_i$; logarithmic differentiation gives the elasticity.
\end{proof}

Let $m$ denote current activity share and $e$ expected activity share on
$A$. Returns $\Pi_A(m),\Pi_B(m)$ are positive and common to users of
the same platform. Platform $A$ also supplies benefit $V(m)+\beta_i e$,
where $\beta_i\geq0$ measures coordination sensitivity. A placement cost
$z_i\in\mathbb R$, independent of $(\bar n_i,\beta_i)$, has continuous
distribution function $F_z$. Comparing the two optimized values gives
the adoption threshold
\begin{align}
 z_i&\leq\Gamma_i(m,e;\varepsilon)
       :=V(m)+\beta_i e+\bar n_iD_\varepsilon(m), \label{act:threshold}\\
 D_\varepsilon(m)&:=
 \frac{\Pi_A(m)^{1+\varepsilon}-\Pi_B(m)^{1+\varepsilon}}
 {1+\varepsilon}.
 \label{act:indirect-difference}
\end{align}
Thus $q_i=F_z(\Gamma_i)$ is the adoption fraction conditional on a type.
Assume $0<\mathbb E[\bar n]<\infty$, and set
\begin{equation}
 A=\mathbb E[\bar n_iq_i],\qquad
 B=\mathbb E[\bar n_i(1-q_i)],\qquad
 \Phi(m,e;\varepsilon)=
 \frac{\Pi_A(m)^\varepsilon A}
 {\Pi_A(m)^\varepsilon A+\Pi_B(m)^\varepsilon B}.
 \label{act:full-share}
\end{equation}
Expectations are integrals over the cross-sectional type distribution.
The continuum specification makes these actual aggregate fractions.
For a finite random population the same ratio of expected activities
need not equal the expectation of the realized activity share.

Headcount adoption is $\mathbb E[q_i]$; activity share is
\eqref{act:full-share}. They measure different objects. For homogeneous
users and a fixed adoption fraction $q\in(0,1)$,
\begin{equation}
 \Phi=\frac{r^\varepsilon q}{r^\varepsilon q+1-q},
 \qquad r=\Pi_A/\Pi_B.
 \label{act:activity-odds}
\end{equation}
At fixed $q$ and $r>1$, activity share increases toward one as
$\varepsilon\to\infty$. This is a conditional allocation comparison:
in the complete choice problem $q$ also changes with deployment value.
It is not a claim of infinite feasible compute or of equal changes in
ownership, voting power and independent decision authority.

\subsection{An exact reduction and its boundaries}
\label{act:reduction}

Write $\ell(x)=(1+e^{-x})^{-1}$ and
$\operatorname{logit}(p)=\log(p/(1-p))$ for $0<p<1$.

\begin{proposition}[From optimal deployment to a logistic share map]
\label{act:exact-reduction}
Suppose users have common $\bar n>0$ and $\beta\geq0$, and
$F_z(z)=\ell((z-\mu)/\tau)$ with $\tau>0$.
Then the activity map satisfies
\begin{equation}
 \operatorname{logit}\Phi(m,e;\varepsilon)
 =\varepsilon\log\frac{\Pi_A(m)}{\Pi_B(m)}
 +\frac{V(m)+\beta e+\bar nD_\varepsilon(m)-\mu}{\tau}.
 \label{act:exact-logit}
\end{equation}
If returns are fixed positive constants and $V(m)=v_0+\alpha m$, then
\begin{align}
 \Phi(m,e;\varepsilon)&=
 \ell\!\left(\frac{\alpha m+\beta e+d(\varepsilon)}{\tau}\right),
 \label{act:reduced-map}\\
 d(\varepsilon)&=v_0-\mu+\bar nD_\varepsilon
           +\tau\varepsilon\log(\Pi_A/\Pi_B).
 \label{act:nonlinear-driver}
\end{align}
\end{proposition}
\begin{proof}
Homogeneity gives $A/B=q/(1-q)$. Taking odds in
\eqref{act:full-share} therefore yields
$\Phi/(1-\Phi)=(\Pi_A/\Pi_B)^\varepsilon q/(1-q)$.
The logistic cost distribution gives
$\operatorname{logit}q=(V+\beta e+\bar nD_\varepsilon-\mu)/\tau$.
The two displays follow by substitution.
\end{proof}

The optimizing model thus admits an exact specialization in which
elasticity moves only an intercept. That intercept is generally nonlinear.
Under the focal-fringe normalization $\Pi_B=1<\Pi_A$,
\begin{equation}
 d'(\varepsilon)=\bar n\int_1^{\Pi_A}x^\varepsilon\log x\,dx
                  +\tau\log\Pi_A>0.
 \label{act:positive-driver}
\end{equation}
Relative advantage alone does not establish this sign at arbitrary return
levels: $D_\varepsilon'=\int_{\Pi_B}^{\Pi_A}x^\varepsilon\log x\,dx$
is negative if both returns lie below one, and its contribution dominates
for sufficiently large $\bar n$. The cost normalization and
absolute returns matter for the adoption value, even when the direct
activity-odds term depends only on their ratio.

State dependence can change geometry rather than merely traverse it.
Under the homogeneous logistic assumptions, differentiating
\eqref{act:exact-logit} at fixed expectations gives
\begin{equation}
 \partial_m\operatorname{logit}\Phi
 =\varepsilon\frac{r'}r+
 \frac{V'+\bar n(\Pi_A^\varepsilon\Pi_A'
                    -\Pi_B^\varepsilon\Pi_B')}{\tau}.
 \label{act:shape-channel}
\end{equation}
Changing $\varepsilon$ now changes the state slope, and potentially its
curvature. Heterogeneous users also generally yield a mixture rather than
one logistic map. A first-order driver approximation is available locally
when $\operatorname{logit}\Phi$ is twice continuously differentiable with
bounded Hessian: its Taylor polynomial has fixed state and expectation
slopes and a linear elasticity term, with a quadratic remainder in the
joint displacement. Such an approximation is local in all those variables.

We henceforth study the reduced family
\begin{equation}
 \Phi_c(m,e)=\ell\!\left(\frac{\alpha m+\beta e+c}{\tau}\right),
 \qquad F_c(m)=\Phi_c(m,m),\qquad G=\alpha+\beta,
 \label{act:family}
\end{equation}
with $\alpha\in\mathbb R$, $\beta\geq0$, $\tau>0$ and real intercept
$c$. Equilibrium expectations satisfy $e=m$. An equilibrium is an
interior fixed point of $F_c$. The nonlinear driver
\eqref{act:nonlinear-driver}, an affine approximation, or an independently
specified institutional intervention can move $c$ through this family.

\subsection{Fixed-point geometry and specified adjustment}
\label{act:geometry}

Define
\begin{equation}
 h(m)=\tau\operatorname{logit}(m)-Gm,
 \qquad v(m)=m(1-m).
 \label{act:graph}
\end{equation}
The equilibrium equation is $h(m)=c$. To attach stability to it, specify
continuous adjustment
\begin{equation}
 \dot m=\kappa\,[F_c(m)-m],\qquad\kappa>0.
 \label{act:adjustment}
\end{equation}
The vector field points into $[0,1]$ at its endpoints, so solutions exist
for all positive times and remain in that interval. Discrete adjustment
$m_{t+1}=F_c(m_t)$ will also be considered where explicitly stated.

\begin{theorem}[Homogeneous equilibrium geometry]
\label{act:geometry-theorem}
The following hold for $\tau>0$.
\begin{enumerate}
\item If $G\leq4\tau$, every $c$ gives exactly one interior equilibrium.
It globally attracts \eqref{act:adjustment}. It is hyperbolic except when
$G=4\tau$, $c=-2\tau$ and $m=1/2$; this exceptional equilibrium remains
attracting.
\item If $G>4\tau$, put
\begin{equation}
 d=\sqrt{1-4\tau/G},\qquad
 m_-=(1-d)/2,\quad m_+=(1+d)/2,
 \qquad c_{\uparrow}=h(m_-),\quad c_{\downarrow}=h(m_+).
 \label{act:folds}
\end{equation}
Then $c_{\downarrow}<c_{\uparrow}$. There are exactly three interior
equilibria when $c_{\downarrow}<c<c_{\uparrow}$, exactly two distinct
interior equilibria at either endpoint, and exactly one outside the
closed interval. In the three-equilibrium case the low and high
equilibria are locally asymptotically stable and the middle one is
unstable under \eqref{act:adjustment}.
\item The same local classification in the three-equilibrium case holds
for discrete adjustment. When $0\leq G\leq4\tau$, its unique
equilibrium globally attracts discrete adjustment as well, including
the exceptional equality case in part 1.
\end{enumerate}
\end{theorem}
\begin{proof}
We have
\begin{equation}
 h'(m)=\frac{\tau}{v(m)}-G,
 \qquad F_c'(m^*)=\frac{Gv(m^*)}{\tau}
 \quad\hbox{at an equilibrium }m^*.
 \label{act:graph-slopes}
\end{equation}
Since $v\leq1/4$, $h$ is strictly increasing when $G\leq4\tau$;
at equality its derivative vanishes only at $1/2$, which does not
destroy strict monotonicity. Its limits at zero and one are $-\infty$
and $+\infty$, proving uniqueness and existence. Moreover
\begin{equation}
 \operatorname{sign}(F_c(m)-m)=\operatorname{sign}(c-h(m)).
 \label{act:drift-sign}
\end{equation}
The continuous flow therefore moves toward the unique root from each
side. Its linearized eigenvalue is
$\kappa(Gv(m^*)/\tau-1)$, strictly negative except in the stated
equality case. The sign argument still proves attraction there.

When $G>4\tau$, $h'$ vanishes exactly at $m_-,m_+$. It is positive
on $(0,m_-)$, negative on $(m_-,m_+)$, and positive on $(m_+,1)$.
These three strictly monotone branches, their endpoint values and
their infinite exterior limits give all the root counts.
On the outer branches $0<F_c'(m^*)<1$, while on the middle branch
$F_c'(m^*)>1$. These inequalities give both local classifications.
At the fold points $h''(m_-)<0<h''(m_+)$, so the double roots are
nondegenerate folds in the intercept parameter.

For the last claim, $G\geq0$ makes $F_c$ nondecreasing. If the first
iterate is above its starting point, all subsequent iterates form a
nondecreasing sequence; otherwise they form a nonincreasing sequence.
They are bounded in $[0,1]$, and continuity makes their limit a fixed
point. Uniqueness identifies it. This argument does not require a strict
linear contraction at the exceptional fixed point.
\end{proof}

For negative $G$, uniqueness and continuous-time attraction do not imply
stability of unrelaxed discrete iteration: a sufficiently negative
$F_c'(m^*)$ can have modulus above one. A stability claim must retain
its adjustment rule. At $G=4\tau,c=-2\tau$, the equality case is also
substantive: writing $m=1/2+y$ gives
$F_c(m)-m=-\tfrac43y^3+O(y^5)$, so the continuous attraction is
nonhyperbolic rather than a strictly negative-eigenvalue conclusion.

\subsection{Response warnings, folds and series poles}
\label{act:warning}

Two derivatives have different meanings. The diagonal derivative
$F_c'=\Phi_m+\Phi_e$ governs the specified adjustment. In contrast,
solving the temporary-equilibrium equation $m=\Phi_c(m,e)$ locally
for $m$ as a function of $e$ gives the response
\begin{equation}
 S=\frac{\Phi_e}{1-\Phi_m}
   =\frac{\beta v(m)}{\tau-\alpha v(m)}.
 \label{act:myopic}
\end{equation}
We restrict attention to the positive-denominator response branch,
$\tau-\alpha v(m)>0$. The implicit derivative itself exists whenever
the denominator is nonzero. Introduce a
nonnegative persistence parameter $\rho$, and, where $|\rho S|<1$,
define the accumulated response
\begin{equation}
 T=S\sum_{j=0}^{\infty}(\rho S)^j
   =\frac{S}{1-\rho S}
   =\frac{\beta v(m)}{\tau-(\alpha+\rho\beta)v(m)}.
 \label{act:warning-index}
\end{equation}
The series describes the sum of a direct response and repeated
expectational echoes with multiplier $\rho S$. It is an additional
response construction; it is not another definition of $F_c'$.

\begin{proposition}[Three distinct thresholds]
\label{act:threshold-separation}
Suppose $\beta>0$ and $0<\rho<1$. On the response domain, $T=1$
occurs at $S=1/(1+\rho)$ and diagonal unit slope occurs at $S=1$;
the latter is the fold condition on the supercritical branches of
Theorem~\ref{act:geometry-theorem}. The response series has its positive
pole at $S=1/\rho$. Consequently
\begin{equation}
 \frac1{1+\rho}<1<\frac1\rho.
 \label{act:threshold-order}
\end{equation}
Writing $H=\alpha+(1+\rho)\beta$, the warning is attainable at an
equilibrium for some intercept if and only if $H\geq4\tau$.
\end{proposition}
\begin{proof}
Solving $S/(1-\rho S)=1$ gives the first value; the series criterion
gives the third. At a homogeneous fold, $Gv=\tau$, so
$\tau-\alpha v=\beta v>0$ and \eqref{act:myopic} gives $S=1$.
The warning equation is $Hv=\tau$. Since $0<v\leq1/4$, it has an
interior solution precisely when $H\geq4\tau$. At such a solution
the myopic denominator is $(1+\rho)\beta v>0$, and
$\rho S=\rho/(1+\rho)<1$, so the solution really lies in the
response domain. Choosing $c=h(m)$ makes it an equilibrium.
\end{proof}

An attainable warning need not imply a fold. With
$(\alpha,\beta,\tau,\rho)=(0,3,1,1/2)$, we have $G=3<4$ but
$H=9/2>4$. At $m=1/3$, take
$c=h(1/3)=-\log2-1$. Then $S=2/3$, $T=1$, and
$F_c'(m)=2/3<1$, although every intercept has a unique equilibrium.
The warning reports accumulated response, not irreversibility.
Likewise, a pole is a failure of the geometric-series representation;
it is not a theorem of finite-time concentration or population explosion.

For $\alpha<4\tau$ and $\rho\geq0$, the warning-attainability boundary
and the critical boundary for the fold regime are
\begin{equation}
 \beta_w=\frac{4\tau-\alpha}{1+\rho},\qquad
 \beta_f=4\tau-\alpha,\qquad
 \beta_f-\beta_w=\frac{(4\tau-\alpha)\rho}{1+\rho}.
 \label{act:coordination-window}
\end{equation}
The warning is attainable at $\beta=\beta_w$, whereas nondegenerate
folds require $\beta>\beta_f$. This compares parameter regions. A statement that a particular evolving
market encounters a warning before a fold requires a path through them.

\subsection{Traversal, warning lead and hysteresis}
\label{act:path}

Assume now $G>4\tau$, $\beta>0$ and $\rho>0$. Let $m_-$ be the
low fold share in \eqref{act:folds}, and define the low warning share
and its intercept by
\begin{equation}
 m_T=\frac{1-\sqrt{1-4\tau/H}}2,
 \qquad c_T=h(m_T).
 \label{act:warning-share}
\end{equation}

\begin{theorem}[Warning before the low fold]
\label{act:path-theorem}
We have $0<m_T<m_-<1/2$ and $c_T<c_{\uparrow}$. The low
equilibrium branch is parameterized by the strictly increasing function
$h:(0,m_-]\to(-\infty,c_{\uparrow}]$.
For an affine channel $c=c_0+\chi\varepsilon$, $\chi>0$, define
\begin{equation}
 \varepsilon_T=\frac{c_T-c_0}{\chi},\qquad
 \varepsilon_f=\frac{c_{\uparrow}-c_0}{\chi}.
 \label{act:path-thresholds}
\end{equation}
Their difference is strictly positive and equals
\begin{equation}
 L:=\varepsilon_f-\varepsilon_T
 =\frac{\tau[\operatorname{logit}(m_-)-\operatorname{logit}(m_T)]
             -G(m_--m_T)}{\chi}.
 \label{act:lead}
\end{equation}
Holding the other parameters fixed, $L$ strictly increases with
$\rho>0$ and tends to zero as $\rho\downarrow0$.
\end{theorem}
\begin{proof}
Since $H=G+\rho\beta>G>4\tau$, the low solution of $Hv=\tau$
lies strictly below the low solution of $Gv=\tau$. The branch
properties follow from the sign of $h'$. In particular,
\begin{equation}
 c_{\uparrow}-c_T
 =\int_{m_T}^{m_-}\left(\frac{\tau}{v(s)}-G\right)ds>0.
 \label{act:lead-integral}
\end{equation}
Integrating and dividing by $\chi$ gives \eqref{act:lead}.
As $\rho$ increases, $H$ increases and $m_T$ strictly decreases,
while $m_-$ stays fixed. Because $h$ is strictly increasing on this
branch, $c_T$ strictly decreases and $L$ strictly increases.
Continuity gives $m_T\to m_-$ and $L\to0$ as $\rho\downarrow0$.
\end{proof}

The path theorem's algebra needs no upper bound on $\rho$. The three
ordered response thresholds in \eqref{act:threshold-order} retain the
stronger condition $\rho<1$. Under that condition the response series
is still convergent at the fold. On the low branch its warning occurs
while the equilibrium remains locally attracting.

The thresholds in \eqref{act:path-thresholds} need not be positive or
accessible. If elasticity is restricted to $I\subset(0,\infty)$, both
thresholds must belong to the allowed path before the comparison describes
a traversal. For a differentiable strictly increasing nonlinear driver
$c=d(\varepsilon)$, the same ordering holds whenever both intercepts
belong to $d(I)$. If $0<l\leq d'\leq u$ between their preimages, then
\begin{equation}
 \frac{c_{\uparrow}-c_T}{u}
 \leq\varepsilon_f-\varepsilon_T
 \leq\frac{c_{\uparrow}-c_T}{l}.
 \label{act:nonlinear-lead}
\end{equation}
This follows by integrating $d'$ and applies directly to admissible
segments of \eqref{act:nonlinear-driver}. A positive driver supplies a
direction, not a guarantee that it reaches either threshold. A lead in
elasticity or intercept units becomes a lead in time only after a law of
motion for that variable is supplied.

\begin{proposition}[Exact hysteresis gap]
\label{act:gap-theorem}
For $G>4\tau$, the two fold intercepts differ by
\begin{equation}
 \Delta:=c_{\uparrow}-c_{\downarrow}
 =Gd-4\tau\operatorname{artanh}(d)>0,
 \qquad d=\sqrt{1-4\tau/G},
 \label{act:gap}
\end{equation}
where $\operatorname{artanh}(d)=\tfrac12\log((1+d)/(1-d))$.
Holding $\alpha$ fixed in the first derivative and $G$ fixed in the second,
\begin{equation}
 \frac{\partial\Delta}{\partial\beta}=d>0,
 \qquad \frac{\partial\Delta}{\partial\tau}
       =-4\operatorname{artanh}(d)<0.
 \label{act:gap-derivatives}
\end{equation}
\end{proposition}
\begin{proof}
Substitute $m_\pm=(1\pm d)/2$ into $h(m_-)-h(m_+)$.
Positivity also follows by integrating $G-\tau/v(s)>0$ between
the fold shares. In differentiating the displayed closed form, all
terms containing a derivative of $d$ cancel because
$4\tau/(1-d^2)=G$. The remaining terms are exactly
\eqref{act:gap-derivatives}.
\end{proof}

For example, $G=6$, $\tau=1$ gives
$m_\pm=(3\pm\sqrt3)/6$,
$c_{\uparrow}=\log(2-\sqrt3)-3+\sqrt3$ and
$c_{\downarrow}=\log(2+\sqrt3)-3-\sqrt3$.
Thus $\Delta=2\sqrt3-4\operatorname{artanh}(1/\sqrt3)>0$.
With $\alpha=4$, $\beta=2$, $\rho=1/2$, the warning gain is $H=7$
and the warning share is $(7-\sqrt{21})/14$, strictly below the low
fold share. Choosing
$\chi=1$ and $c_0=c_T-1$ makes the warning elasticity exactly one
and the fold elasticity $1+c_{\uparrow}-c_T>1$, an explicit
positive-elasticity path through both thresholds.

\begin{figure}[tbp]
\centering
\begin{minipage}{0.49\textwidth}
\centering
\begin{tikzpicture}
\begin{axis}[
 width=\linewidth,height=6.4cm,
 xmin=-4.2,xmax=-1.8,ymin=0,ymax=1,
 xlabel={Intercept $c$},ylabel={Activity share $m$},
 title={Equilibrium branches},
 tick label style={font=\footnotesize},label style={font=\small},
 title style={font=\small},grid=major,grid style={gray!15},
 xtick={-4,-3.5,-3,-2.5,-2},ytick={0,0.25,0.5,0.75,1}]
\addplot[inkblue,very thick,domain=0.005:0.2113248654,samples=100]
 ({ln(x/(1-x))-6*x},{x});
\addplot[gray!75!black,thick,dashed,domain=0.2113248654:0.7886751346,samples=100]
 ({ln(x/(1-x))-6*x},{x});
\addplot[inkblue,very thick,domain=0.7886751346:0.995,samples=100]
 ({ln(x/(1-x))-6*x},{x});
\addplot[only marks,mark=*,mark size=1.8pt,black]
 coordinates {(-2.5849070894,0.2113248654) (-3.4150929106,0.7886751346)};
\node[font=\scriptsize,anchor=west] at (axis cs:-2.57,0.29) {Low fold};
\node[font=\scriptsize,anchor=east] at (axis cs:-3.44,0.71) {High fold};
\end{axis}
\end{tikzpicture}
\end{minipage}\hfill
\begin{minipage}{0.49\textwidth}
\centering
\begin{tikzpicture}
\begin{axis}[
 width=\linewidth,height=6.4cm,
 xmin=-2.72,xmax=-2.575,ymin=0.11,ymax=0.235,
 xlabel={Intercept $c$},ylabel={Activity share $m$},
 title={Low branch: warning before fold},
 tick label style={font=\footnotesize},label style={font=\small},
 title style={font=\small},grid=major,grid style={gray!15},
 xtick={-2.7,-2.65,-2.6},ytick={0.12,0.15,0.18,0.21},
 scaled x ticks=false,xticklabel style={/pgf/number format/fixed}]
\addplot[inkblue,very thick,domain=0.11:0.2113248654,samples=100]
 ({ln(x/(1-x))-6*x},{x});
\addplot[gray!75!black,thick,dashed,domain=0.2113248654:0.235,samples=50]
 ({ln(x/(1-x))-6*x},{x});
\addplot[only marks,mark=*,mark size=2pt,orange!80!black]
 coordinates {(-2.6028382248,0.1726731646)};
\addplot[only marks,mark=*,mark size=1.8pt,black]
 coordinates {(-2.5849070894,0.2113248654)};
\node[font=\scriptsize,anchor=east,text=orange!70!black]
 at (axis cs:-2.62,0.176) {$T=1$};
\node[font=\scriptsize,anchor=east]
 at (axis cs:-2.602,0.218) {$F_c'=1$};
\end{axis}
\end{tikzpicture}
\end{minipage}
\caption{The analytical equilibrium curve $c=\log(m/(1-m))-6m$ for
$\alpha=4$, $\beta=2$, $\tau=1$, and $\rho=1/2$. Away from the folds,
solid branches are locally attracting under~\eqref{act:adjustment}; the dashed branch is
unstable. The right panel enlarges the low branch. Its accumulated-response
warning occurs at $m_T=(7-\sqrt{21})/14$, before the fold at
$m_-=(3-\sqrt3)/6$. These are evaluations of the exact model curve,
not simulated or fitted observations.}
\label{fig:act-geometry}
\end{figure}

In a quasi-static branch-following interpretation, a population initially
on the low branch loses that branch as $c$ increases past
$c_{\uparrow}$. A population on the high branch retains it until $c$
decreases past $c_{\downarrow}$. The gap is the additional intercept
reduction associated with eliminating the high branch rather than
preserving the low one. It is not a monetary policy cost, and it is not
the minimum finite-time intervention: a temporary intervention that moves
the state across the middle equilibrium can change its basin without
removing the high branch. Finite-rate branch tracking and delayed tipping
require their own dynamic analysis.

\subsection{Heterogeneity and influence at the margin}
\label{act:heterogeneity}

Now consider finitely many user types with masses $\mu_i\geq0$,
$\sum_i\mu_i=1$, activity weights $n_i\geq0$ and
$N=\sum_i\mu_i n_i>0$. These weights are fixed within the following
share map; allowing them to respond to the state reintroduces the
additional channels in \eqref{act:full-share}. With $\tau>0$ and
coordination coefficients $\beta_i\geq0$, define
\begin{equation}
 \Phi^w(m,e)=\frac{\mathbb E[n_i
       \ell((\alpha m+\beta_i e+c)/\tau)]}{N},
 \quad p_i(m)=\ell(((\alpha+\beta_i)m+c)/\tau),
 \quad F^w(m)=\Phi^w(m,m).
 \label{act:mixture}
\end{equation}
This is a further reduced specification: heterogeneous opportunity scales
in \eqref{act:threshold} generally also create type-specific intercepts,
which the common $c$ in \eqref{act:mixture} does not retain.
The expectation now uses the finite law $\mu$. At a state $m$, put
\begin{equation}
 V=\mathbb E[n_i v(p_i)],\qquad
 B=\mathbb E[n_i\beta_i v(p_i)],\qquad
 \beta^{\rm marg}=B/V.
 \label{act:marginal-quantities}
\end{equation}
Finite logits and $N>0$ ensure $V>0$.

\begin{proposition}[Marginal weights and response loads]
\label{act:marginal-theorem}
The exact derivatives and covariance identity are
\begin{align}
 (F^w)'&=\frac{\alpha V+B}{\tau N},&
 \beta^{\rm marg}-\mathbb E[\beta]
 &=\frac{\operatorname{Cov}(\beta,nv(p))}{V}.
 \label{act:marginal-formulas}
\end{align}
At an equilibrium where $\tau N-\alpha V>0$, the implicit response is
$S=B/(\tau N-\alpha V)$. Where also $|\rho S|<1$,
\begin{equation}
 T=\frac{B}{\tau N-\alpha V-\rho B}.
 \label{act:hetero-index}
\end{equation}
On this domain, $T=1$ is equivalent to
$\alpha V+(1+\rho)B=\tau N$, whereas $(F^w)'=1$ is equivalent
to $\alpha V+B=\tau N$. Their load difference is $\rho B$.
\end{proposition}
\begin{proof}
Differentiate each logistic term: its own-state partial is
$\alpha v(p_i)/\tau$ and its expectation partial is
$\beta_i v(p_i)/\tau$. Summation yields the derivative and implicit
response. Summing the geometric series yields
\eqref{act:hetero-index}. The covariance identity is the elementary
equality $\mathbb E[\beta w]-\mathbb E[\beta]\mathbb E[w]
=\operatorname{Cov}(\beta,w)$ with $w=nv(p)$.
The load equivalences follow by multiplying the nonzero denominators.
\end{proof}

The users with the largest activity-weighted susceptibility combine
substantial activity with substantial density near the adoption threshold.
Here $v(p_i)/\tau$ measures response to a small common-index shift;
it is not a temporal switching probability. Retaining a user's latent
cost and redrawing that cost each period can produce the same static
adoption probability but different switching rates. A user already very
likely to choose one platform has small $v(p_i)$, even if its activity is large.
For comparison, pure activity weighting gives
$\mathbb E[n\beta]/N-\mathbb E[\beta]
=\operatorname{Cov}(\beta,n)/N$.
The two covariances can have opposite signs.

\begin{proposition}[A reversal between activity and susceptibility weighting]
\label{act:covariance-reversal}
Take two equally numerous types with
$\alpha=0$, $\tau=1$, $c=0$, and evaluate at $m=1/2$. Set
\begin{equation}
 (\beta_L,\beta_H)=(0,2\log3),\qquad
 (n_L,n_H)=(1,6/5).
 \label{act:reversal-parameters}
\end{equation}
Then
\begin{align}
 \operatorname{Cov}(\beta,n)&=\frac{\log3}{10}>0,&
 \operatorname{Cov}(\beta,nv(p))&=-\frac{\log3}{80}<0,\label{act:reversal-cov}\\
 \beta^{\rm activity}&=\frac{12}{11}\log3
   >\mathbb E[\beta]=\log3
   >\beta^{\rm marg}=\frac{18}{19}\log3.
 \label{act:reversal-means}
\end{align}
\end{proposition}
\begin{proof}
The adoption probabilities are $(1/2,3/4)$ and the marginal weights
are $(1/4,9/40)$. For two equal masses,
$\operatorname{Cov}(a,b)=(a_H-a_L)(b_H-b_L)/4$.
Substitution gives the covariance values; dividing the corresponding
weighted sums gives the three means.
\end{proof}

This is a statewise witness, not an equilibrium at $m=1/2$:
$F^w(1/2)=7/11$. It establishes exactly why an activity statistic
cannot substitute for the derivative governing adjustment. More
precisely, it reverses the position of a weighted mean coordination
coefficient relative to its unweighted mean. It does not by itself
reverse individual response contributions or imply a smaller adjustment
slope. In this example $(F^w)'(1/2)=9\log3/44$, whereas equal activity
weights give $3\log3/16$: the former is $12/11$ times the latter.
More generally, for two equal masses, positive activities and
$\beta_H>\beta_L$, the same sign reversal occurs precisely when
\begin{equation}
 1<\frac{n_H}{n_L}<\frac{v(p_L)}{v(p_H)}.
 \label{act:reversal-region}
\end{equation}

If $(F^w)'<1$ everywhere in $(0,1)$, the map has exactly one interior
fixed point: $F^w(m)-m$ strictly decreases from a positive to a negative
endpoint value. Conversely, two distinct fixed points force an
intermediate point with derivative one by Rolle's theorem. A convenient
sufficient condition, when every $\alpha+\beta_i\geq0$, is
$\mathbb E[n_i(\alpha+\beta_i)]<4\tau N$, since $v(p_i)\leq1/4$.
The reversed inequality is not a theorem of multiple equilibria.
Neither is slope one alone a fold; nondegeneracy and parameter
transversality are additional requirements. The statewise load separation
in Proposition~\ref{act:marginal-theorem} consequently does not supply a
constant warning window or a heterogeneous warning-before-fold path theorem.

\subsection{A heterogeneous concentration staircase}
\label{act:staircase}

Heterogeneity can create more than the homogeneous model's three
equilibria. The following witness gives both exact inequalities and a
dynamic consequence \cite{VallierWCCAppendix}.

\begin{theorem}[Five fixed-point intervals and three invariant bands]
\label{act:five-roots}
With two equal masses and activities, take
$\alpha=0$, $\tau=1$, $c=-14\log3$ and
$(\beta_1,\beta_2)=(20\log3,40\log3)$. The mixture map is
\begin{equation}
 F(m)=\frac12\left[
 \ell((20m-14)\log3)+\ell((40m-14)\log3)\right].
 \label{act:five-map}
\end{equation}
It has at least one fixed point in each of
\begin{equation}
 (0,1/4),\quad(1/4,9/20),\quad(9/20,11/20),\quad
 (11/20,4/5),\quad(4/5,1).
 \label{act:five-intervals}
\end{equation}
For discrete adjustment $m_{t+1}=F(m_t)$, every orbit in $[0,1]$
converges to an interior fixed point. The three disjoint bands
\begin{equation}
 [0,1/4],\qquad[9/20,11/20],\qquad[4/5,1]
 \label{act:three-bands}
\end{equation}
are forward invariant, and an orbit starting in one of them converges
to a fixed point strictly inside that band's endpoints. Thus starts in
the three different bands have three necessarily distinct limits.
\end{theorem}
\begin{proof}
The map is continuous, strictly increasing, and takes values in $(0,1)$.
Using $\ell(k\log3)=3^k/(1+3^k)$ gives the following bounds:
\begin{equation}
 \begin{array}{c|cccccc}
 m&0&1/4&9/20&11/20&4/5&1\\ \hline
 \operatorname{sign}(F(m)-m)&+&-&+&-&+&-
 \end{array}
 \label{act:five-signs}
\end{equation}
Indeed, $F(1/4)<1/82<1/4$ because its two exponents are $-9,-4$.
At $9/20$ the exponents are $-5,4$, so
$F(9/20)>81/164>9/20$.
At $11/20$ they are $-3,8$, so
$F(11/20)<(1/28+1)/2=29/56<11/20$.
At $4/5$ they are $2,18$, so $F(4/5)>9/10>4/5$.
The endpoint signs follow from $0<F<1$. The intermediate value
theorem gives a root in each open interval in \eqref{act:five-intervals}.

For each band $[a,b]$ in \eqref{act:three-bands}, the endpoint signs
give $F(a)>a$ and $F(b)<b$. Monotonicity implies
$F([a,b])\subset(a,b)$, proving invariance. Every orbit is monotone
after comparing its first two terms and is bounded; its limit is a
fixed point by continuity. Strict endpoint inequalities exclude a
band endpoint as the limit. The same monotone-orbit argument on
$[0,1]$ proves convergence from every initial state, and $F(0)>0$,
$F(1)<1$ exclude the exterior endpoints.
\end{proof}

The theorem does not assert that there are exactly five roots, that
each band contains only one, or that the displayed band endpoints are
basin boundaries. Its positive conclusion is already substantial:
different groups' adoption margins support separated long-run
concentration ranges. Replacing the mixture by a logistic map with a
single average coordination parameter would remove this structure.

\subsection{Institutional channels and the meaning of concentration}
\label{act:institutions}

An institution can change the deployment returns rather than the
intercept alone. For example, suppose the leading platform provides
return $\Pi_A(m)=\Pi_0(m)-\chi_qW(\Psi(m),k(m))$, where
$\chi_q>0$, $W_\Psi>0$ and $W_k<0$. Here $\Psi$ measures load and
$k$ service priority. Then
\begin{equation}
 \Pi_A'(m)=\Pi_0'(m)+
 \chi_q\big[(-W_k)k'(m)-W_\Psi\Psi'(m)\big].
 \label{act:queue-channel}
\end{equation}
Priority advantages and congestion work in opposite directions. A
priority-neutral rule removes the $k'$ channel; a capacity investment
can change load; an interoperability rule can change placement costs,
coordination benefits or both. Through \eqref{act:shape-channel}, these
interventions can change the equilibrium geometry itself. Assigning all
of them to one intercept would require a further approximation.

The gap derivatives in \eqref{act:gap-derivatives} state what happens
when coordination or dispersion changes while the other specified
primitives are held fixed. They do not determine which policy produces
that change or whether the intervention improves welfare. Productive
coordination and imposed switching losses can both encourage adoption;
their welfare consequences differ. A welfare analysis must specify real
benefits, resource costs, external effects and who receives them.
Likewise, concentration of activity does not alone establish ownership
concentration, domination or misalignment.

The application connects evaluated deployment to a changing aggregate
environment without assuming that concentration is universally desirable
or unavoidable. Its main deductions concern how the resulting equations
behave: when there is one equilibrium or several, how warnings differ
from folds, how a deployment channel traverses those thresholds, and how
heterogeneity changes the possible long-run states. Reproductive
transmission can then be added to determine how the decision procedures
and mandates selecting those deployments persist across actual
successors. The economic and evolutionary mechanisms meet at that
explicit interface.

\clearpage
\section{Endogenous electorates and the reproduction of influence}
\label{app:elect}

Replication can change who participates in a collective decision. An agent
that creates a second instance may acquire a second vote, divide an existing
vote between its instances, or create an instance with no voting rights.
These are different institutions applied to the same copying technology.
The distinction matters even when the copies inherit exactly the same
preferences and submit sincere ballots. If each recognized copy receives
one vote, choosing a population becomes a way of choosing the outcome of
an election.

This appendix develops that application in three stages. First, a spawning
game connects outcome preferences and copying costs to the composition of
an electorate. Second, exact impossibility and possibility results identify
what equal votes, conserved influence, and a protected status quo permit.
Finally, a fully specified population dynamic shows how electoral outcomes
can feed back into reproduction. The static results draw on
\cite{VallierElectorate,VallierCFS}; the final dynamic specializes the
winner-feedback mechanism studied in \cite{VallierMI}. False-name voting
and costly identity creation provide the relevant antecedents
\cite{Conitzer2008,WagmanConitzer2008}. The copying operation here is
restricted: additional ballots reproduce the chooser's sincere ranking.

\subsection{Sincere types and strategic copy counts}
\label{elect:game}

Let $A$ be a finite set of alternatives with $|A|\geq2$, and let
$\mathcal L(A)$ be its set of strict linear orders. A profile $P$ is a
nonempty finite multiset of orders. Its size is $|P|$, with repetitions
counted. A resolute voting rule $f$ assigns one alternative to every such
profile. For $\sigma\in\mathcal L(A)$ and integer $k\geq0$, write
$P\oplus k\sigma$ for the profile obtained by adding $k$ copies of
$\sigma$. This operation models copies that are admitted as additional
equal-weight voters.

There is a finite set $T$ of strategic types. Type $\theta$ has a sincere
order $\sigma_\theta$, an outcome utility $u_\theta:A\to\mathbb R$, and
an initial count $n_\theta\geq0$, where
$\sum_\theta n_\theta\geq1$. Utilities represent the orders exactly:
$a\succ_{\sigma_\theta}b$ if and only if
$u_\theta(a)>u_\theta(b)$. A type with zero initial count is a potential
entrant. To study only replication of existing types, restrict the
strategic set to positive-count types.

Each type chooses an integer $k_\theta\geq0$ and pays
$C_\theta(k_\theta)$. Assume $C_\theta(0)=0$ and that $C_\theta$ is
nonnegative and nondecreasing. The resulting profile and payoff are
\begin{equation}
 P(k)=\biguplus_{\theta\in T}(n_\theta+k_\theta)\sigma_\theta,
 \qquad
 U_\theta(k)=u_\theta(f(P(k)))-C_\theta(k_\theta).
 \label{elect:payoff}
\end{equation}
Costs are expressed in the same payoff units as $u_\theta$. Rescaling
utility while holding these numerical costs fixed would change the game.
The preferences submitted by a type are fixed; its strategic choice is
how many recognized copies to create.

\begin{definition}[Cost-adjusted population stability]
\label{elect:cost_stability}
Writing $P_0=P(0)$, the initial electorate is cost-adjusted population
stable if, for every $\theta\in T$ and every integer $k\geq0$,
\begin{equation}
 u_\theta(f(P_0\oplus k\sigma_\theta))-u_\theta(f(P_0))
 \leq C_\theta(k).
 \label{elect:stability_inequality}
\end{equation}
\end{definition}

\begin{proposition}[Zero-spawning equilibrium]
\label{elect:zero_nash}
The all-zero strategy profile is a pure Nash equilibrium of
\eqref{elect:payoff} if and only if
\eqref{elect:stability_inequality} holds.
\end{proposition}

\begin{proof}
A unilateral deviation from the all-zero profile changes the electorate
from $P_0$ to $P_0\oplus k\sigma_\theta$. Since $C_\theta(0)=0$,
the deviator's payoff change is precisely the left-hand side of
\eqref{elect:stability_inequality} minus $C_\theta(k)$. All these
changes are nonpositive exactly when zero spawning is Nash.
\end{proof}

The equivalence supplies the strategic interpretation of stability in
this particular game. It does not establish the existence of another
equilibrium when zero spawning fails. Nor does it introduce a new
independent decision maker for every copy: the controllers in $T$ are
fixed, while their represented populations are endogenous. Autonomous
descendants with distinct objectives and continuation strategies require
the additional ownership and information structure discussed elsewhere
in the paper.

There is a stronger, useful deterrence criterion. For a fixed type and
positive count $k$, let $G_\theta(k)$ be the largest outcome-utility
gain from adding those copies over all opponents' count choices. This
largest gain exists: the opponents have at least their zero choice, and
every gain belongs to the finite set
$\{u_\theta(a)-u_\theta(b):a,b\in A\}$. Zero spawning is a weakly
dominant strategy exactly when $C_\theta(k)\geq G_\theta(k)$ for
every $k\geq1$. Indeed, these are exactly the unilateral comparisons
against every opponents' strategy. In particular, writing
\[
 R_\theta=\max_{a\in A}u_\theta(a)-\min_{a\in A}u_\theta(a),
\]
$C_\theta(1)\geq R_\theta$ suffices; a strict inequality makes zero
strictly preferable to every positive count. Monotonicity of costs and
the bound on outcome gains prove both assertions. Institutional
susceptibility to extra votes can therefore coexist with economic
deterrence of their creation.

\subsection{One existing copy can defeat population stability}
\label{elect:obstruction}

For distinct alternatives $a,b$, let $N_{ab}(P)$ count voters ranking
$a$ above $b$, and define the integer margin
$M_{ab}(P)=N_{ab}(P)-N_{ba}(P)$. An alternative $c$ is a strict
Condorcet winner if $M_{cb}(P)>0$ for every $b\ne c$. It is a weak
Condorcet winner if all those inequalities are weak. A rule is
Condorcet-consistent if it selects every strict Condorcet winner.
There is at most one such winner, since two distinct winners would have
to beat each other.

Call a rule restricted one-copy population stable if, for every profile
$P$ and every ballot $\sigma$ already present in $P$,
\begin{equation}
 f(P\oplus\sigma)\not\succ_\sigma f(P).
 \label{elect:ordinal_stability}
\end{equation}
This is a cost-free condition on the outcome comparison. It asks for
protection even against one faithful copy of an existing voter.

\begin{theorem}[One-copy Condorcet obstruction]
\label{elect:one_copy}
No resolute Condorcet-consistent rule on all nonempty finite profiles
satisfies \eqref{elect:ordinal_stability} when $|A|\geq2$.
\end{theorem}

\begin{proof}
Choose distinct $a,c\in A$ and fix a common order of the remaining
alternatives. Let $\sigma_a$ rank $a$ first and $c$ second, and let
$\sigma_c$ rank $c$ first and $a$ second, with the common remaining
order below them. If there are only two alternatives, the lower tail is
empty. Take $P$ to contain two copies of each ballot.

The margin between $a$ and $c$ is zero, and both beat every outsider
unanimously. If $f(P)=a$, add one $\sigma_c$. Then $c$ beats $a$ by
three votes to two and beats every outsider unanimously. Condorcet
consistency forces the new outcome to be $c$, a strict improvement for
the existing $\sigma_c$ type. If $f(P)=c$, the symmetric addition of
$\sigma_a$ improves that type's outcome. If $f(P)$ is an outsider,
adding $\sigma_a$ makes $a$ the strict Condorcet winner and again
strictly improves the copying type's outcome. Every possible output
therefore violates \eqref{elect:ordinal_stability}.
\end{proof}

The theorem concerns every rule but only some profile for each rule.
It neither says that every tied election is manipulable by one copy nor
that the improvement exceeds a positive copying cost. Its force is
institutional: sincere transmission, existing-type replication, and
equal recognition already suffice. Detecting that a voter is a copy
does not change the proof if the institution still admits its ballot
at full weight. Excluding copies or conserving their total influence
changes the institution to which the theorem applies.

\subsection{Forcing numbers, stakes, and copying costs}
\label{elect:forcing_costs}

The one-copy witness concerns a particular boundary. For an arbitrary
profile, the number of copies needed to install a strict Condorcet
winner can be computed exactly. For any ballot $\sigma$ and distinct
$c,b$, the elementary update is
\begin{equation}
 M_{cb}(P\oplus k\sigma)=
 M_{cb}(P)+k
 \begin{cases}
  1,&c\succ_\sigma b,\\
  -1,&b\succ_\sigma c.
 \end{cases}
 \label{elect:margin_update}
\end{equation}
Thus every comparison matters. A copy that places $c$ below an opponent
lowers that margin. Copies placing $c$ first raise every relevant margin
simultaneously.

\begin{theorem}[Exact top-ballot forcing number]
\label{elect:forcing}
Let $|P|=n\geq1$, fix $c\in A$, and let $\sigma$ rank $c$ first.
The least nonnegative integer $k$ making $c$ a strict Condorcet winner
of $P\oplus k\sigma$ is
\begin{equation}
 k^*(P,c)=\max\left\{0,\,1-\min_{b\ne c}M_{cb}(P)\right\}.
 \label{elect:kstar}
\end{equation}
It is independent of the ordering below $c$, satisfies
$0\leq k^*(P,c)\leq n+1$, and is zero exactly when $c$ is already a
strict Condorcet winner. If every original voter ranks $c$ last, then
$k^*(P,c)=n+1$.
\end{theorem}

\begin{proof}
Equation \eqref{elect:margin_update} gives
$M_{cb}(P\oplus k\sigma)=M_{cb}(P)+k$. Since margins and $k$ are
integers, this is positive if and only if
$k\geq1-M_{cb}(P)$. Satisfying these inequalities for all opponents,
together with $k\geq0$, gives exactly \eqref{elect:kstar}. An opponent
attaining the minimum margin exists because $A\setminus\{c\}$ is
nonempty and finite, so any smaller nonnegative count fails.
Every original margin is at least $-n$, giving the upper bound. Zero
copies suffice exactly when every original integer margin is at least
one. If all voters put $c$ last, each margin is $-n$ and the formula
gives $n+1$.
\end{proof}

Every Condorcet extension must select $c$ once this threshold is reached.
The threshold is therefore independent of its tie-breaking and
cycle-resolution choices. It measures copies sufficient for this
particular guarantee. A specified voting rule may select $c$ earlier,
before $c$ defeats every opponent.

\begin{proposition}[Deterring a forcing move]
\label{elect:forcing_cost}
Suppose $f$ is Condorcet-consistent and a present type $\theta$ has
top alternative $c\ne f(P)$. Write
$B_\theta=u_\theta(c)-u_\theta(f(P))>0$. Among moves that add
enough copies of $\sigma_\theta$ to make $c$ a strict Condorcet
winner, minimal forcing maximizes net benefit. A profitable such move
exists if and only if
\begin{equation}
 B_\theta>C_\theta(k^*(P,c)).
 \label{elect:profitable_forcing}
\end{equation}
For linear cost $C_\theta(k)=q_\theta k$, all these forcing moves
are deterred exactly when
$q_\theta\geq B_\theta/k^*(P,c)$.
\end{proposition}

\begin{proof}
Because $f(P)\ne c$, Condorcet consistency implies that $c$ is not
already a strict winner, so $k^*(P,c)\geq1$. Every count
$k\geq k^*(P,c)$ gives the same outcome $c$ and the same gross
benefit $B_\theta$. Nondecreasing cost makes the smallest such count
weakly cheapest. Its net benefit is positive exactly under
\eqref{elect:profitable_forcing}. Dividing the linear-cost inequality
by the positive forcing count proves the final assertion.
\end{proof}

Types already receiving their top alternative have no positive gross
gain available and need no ratio with a possibly zero forcing count.
For the other types, the ratio separates the value of changing the
outcome from the number of recognized copies needed to guarantee it.
This is the useful cost characterization in \cite{VallierCFS}.

\paragraph{A cost comparison at a tied profile.}
Consider four voters, two ranking $a\succ b\succ d$ and two ranking
$b\succ a\succ d$. Choose a Condorcet extension returning $a$ at
this tie, and give the second type utilities
$(u(b),u(a),u(d))=(1,1/2,0)$. Its forcing count is one and its gross
gain is $1/2$. Linear per-copy cost $q$ deters forcing exactly when
$q\geq1/2$. In particular, $q=1/(|P|+1)=1/5$ leaves net benefit
$3/10$. The universal upper bound $k^*\leq n+1$ does not imply that
per-copy cost $1/(n+1)$ deters all unit-range gains. For utilities in
$[0,1]$, cost at least one per copy is a uniform sufficient bound;
smaller sufficient costs depend on the particular stakes and margins.

\paragraph{Deterring forcing need not deter every manipulation.}
Take two alternatives, two $a$-first voters and one $b$-first voter.
Use majority rule with ties resolved as $b$. A $b$-type with
$u(b)=1$, $u(a)=0$ and per-copy cost $3/4$ needs two copies to make
$b$ a strict Condorcet winner, so strict-winner forcing is unprofitable.
One copy, however, creates a tie resolved as $b$ and yields net benefit
$1/4$. The cost criterion in Proposition~\ref{elect:forcing_cost}
therefore concerns the specified forcing class. The full Nash condition
remains \eqref{elect:stability_inequality}, which checks every count
against the actual rule.

\subsection{Conserving influence while allowing replication}
\label{elect:weights}

The preceding results assign a new unit of influence to each admitted
copy. A different constitution can let carriers multiply while their
collective voting weight stays fixed. Let a weighted profile be a finite
list of pairs $(\sigma_i,s_i)$, with $s_i>0$ and
$\sum_i s_i=1$. Define its aggregate ballot weights and weighted
margins by
\[
 w_P(\sigma)=\sum_{i:\sigma_i=\sigma}s_i,
 \qquad
 M^w_{ab}(P)=\sum_{i:a\succ_{\sigma_i}b}s_i
              -\sum_{i:b\succ_{\sigma_i}a}s_i.
\]
Consider a complete rule of the form $F(P)=\widehat F(w_P)$, where
$\widehat F$ selects the weighted Condorcet winner whenever one exists.
Such a rule exists: select the unique weighted winner when present and
otherwise return a fixed default alternative. More generally, the
fallback can be any specified function of $w_P$.

\begin{theorem}[Conserved-weight splitting]
\label{elect:split}
Replace any voter $(\sigma,s)$ by $k\geq1$ same-ballot copies with
positive weights $s_1,\ldots,s_k$ satisfying $\sum_j s_j=s$.
Every aggregate ballot weight, every weighted margin, and the outcome
of $F$ remain unchanged. Any finite succession of these operations has
the same property.
\end{theorem}

\begin{proof}
The replacement removes weight $s$ from ballot $\sigma$ and adds
$\sum_j s_j=s$ to that same ballot. Every other aggregate ballot
weight is untouched. Thus $w_P$ is unchanged. Weighted margins are
functions of $w_P$, as is the complete rule $F$, so both are unchanged.
Induction gives the assertion for finitely many splits.
\end{proof}

The requirement on the entire rule is essential. Preserving margins
alone does not preserve the outcome of a fallback that consults the
number of identities. For example, at a weighted tie a rule could
choose different defaults for even and odd identity counts while
remaining weighted-Condorcet-consistent whenever a strict winner
exists. Splitting could then change its fallback outcome.

In a game where each controller's only permitted modification divides
a fixed stock of ballot weight among faithful copies, every joint
splitting profile has the same outcome. Nonnegative splitting costs
make no split weakly dominant; a strictly positive cost for every
nontrivial split makes it strictly preferable to those splits. This
conclusion does not require a particular outcome utility. Its operative
assumptions are conserved ballot influence and a rule that depends on
that influence. Acquiring new weight or changing the ballots is a
different admissible action. Replication can also remain valuable for
production even when it yields no extra electoral power.

\subsection{A protected status quo under bounded penetration}
\label{elect:supermajority}

A second construction permits additional equal-weight identities but
limits their share and sacrifices responsiveness. Following the
status-quo approach to sybil resilience
\cite{ShahafShapiroTalmon2019}, fix a distinguished alternative $r$.
The rule below uses a supermajority against every opponent. Its proof
is direct; it does not depend on identifying it with a rule that first
filters a set of viable alternatives.

Partition the observed electorate as $V=H\uplus S$, where $H$ is a
nonempty genuine profile and $S$ is a profile of additional identities
whose ballots may be arbitrary. Put $h=|H|$, $s=|S|$, and $v=h+s$.
Fix a known bound $0\leq\alpha<1$ and assume $s/v\leq\alpha$.
The rule need not identify the partition; the guarantee is conditional
on its actual penetration satisfying this bound.

For any nonempty profile $Q$, let
\[
 C_r(Q)=\{a\ne r:N_{ar}(Q)>|Q|/2\}.
\]
Define a genuine-voter base rule $R_0(Q)$ as follows: if the majority
relation restricted to $C_r(Q)$ has a strict Condorcet winner, select
it; otherwise select $r$. The empty set yields $r$, and a singleton
set has its sole member as its winner. Define the observed-electorate
rule $R_\alpha(V)$ to select $c\ne r$ if
\begin{equation}
 N_{cb}(V)>\frac{1+\alpha}{2}v
 \quad\hbox{for every }b\ne c;
 \label{elect:full_pair}
\end{equation}
if there is no such $c$, return $r$. There cannot be two such
candidates, since each would have to defeat the other with more than
half the votes.

\begin{theorem}[Full-pair safety and liveness]
\label{elect:safety}
For every partition satisfying $s/v\leq\alpha$, a non-status-quo
output of $R_\alpha(V)$ equals $R_0(H)$. If $\alpha<1/3$, then,
for every fixed sybil profile and every target $c\ne r$, the genuine
voters can choose ballots causing $R_\alpha$ to elect $c$. The latter
uniform guarantee cannot include the endpoint $\alpha=1/3$.
\end{theorem}

\begin{proof}
Suppose the output is $c\ne r$. For each opponent $b$, at most $s$
supporters of $c$ over $b$ belong to $S$. Consequently
\[
 N_{cb}(H)\geq N_{cb}(V)-s
 >\frac{1+\alpha}{2}v-s
 \geq\frac{v-s}{2}=\frac h2,
\]
where the last inequality is $s\leq\alpha v$. Thus $c$ is the
strict Condorcet winner among genuine voters on all of $A$. In
particular, $c\in C_r(H)$ and defeats every other member of that
set, so $R_0(H)=c$.

For liveness, hold the sybil ballots fixed and let every genuine voter
rank the target $c$ first. For every $b\ne c$,
\[
 N_{cb}(V)\geq h\geq(1-\alpha)v
 >\frac{1+\alpha}{2}v
\]
when $\alpha<1/3$. Hence \eqref{elect:full_pair} holds. At the
endpoint, take two alternatives, two genuine voters, and one sybil
ranking $r$ first. Even unanimous genuine support for $c$ supplies
only two of the three votes, exactly the threshold when
$\alpha=1/3$. No genuine ballot choice can exceed it.
\end{proof}

The safety statement also implies that if $R_0(H)=r$, the combined
rule returns $r$. It permits an ordinary genuine majority to be
insufficient for change. This is the price of its protection. The
notion of safety is explicitly relative to the genuine-voter base;
it does not identify the genuine outcome with welfare or alignment.
The penetration bound must also hold after the extra identities have
been admitted. Merely announcing a bound cannot make it true.

\subsection{When electoral victory changes reproduction}
\label{elect:dynamic}

The spawning game gives reproduction a political payoff. To close a
dynamic feedback loop, specify how a political outcome changes future
reproduction. The following two-type model isolates the mechanism in
\cite{VallierMI} without requiring a global description of every
multi-alternative switching surface.

There are two inherited rankings, $a\succ b$ and $b\succ a$. Let
$x\in[0,1]$ be the population share of the first type. Simple majority
selects $a$ when $x>1/2$ and $b$ when $x<1/2$. Under outcome $a$,
the two types' per-capita growth rates are $(r+\kappa,r)$; under
outcome $b$, they are $(r,r+\kappa)$, where $r>0$ and
$\kappa>0$. This is a specified resource-to-reproduction advantage
for the winning preference type. Growth rates are not defined as its
outcome utility.

Away from a tie, normalizing the two absolute populations gives
$\dot x=x(1-x)(w_a-w_b)$. At the discontinuity, use the explicit
convexified inclusion
\begin{equation}
 \dot x\in\mathcal F(x):=
 \begin{cases}
 \{-\kappa x(1-x)\},&0\leq x<1/2,\\
 [-\kappa/4,\kappa/4],&x=1/2,\\
 \{\kappa x(1-x)\},&1/2<x\leq1.
 \end{cases}
 \label{elect:inclusion}
\end{equation}
A solution is an absolutely continuous path in $[0,1]$ satisfying
this inclusion almost everywhere. The middle interval is exactly the
convex hull of the two one-sided limiting velocities. Equation
\eqref{elect:inclusion} fully specifies the model, including the
tie; an arbitrary tie-breaking value is not being treated as a smooth
vector field.

\begin{proposition}[Majority reinforcement]
\label{elect:repulsion}
Every solution starting with $0<x(0)<1/2$ converges to zero, and every
solution starting with $1/2<x(0)<1$ converges to one. Both endpoints
are asymptotically stable relative to $[0,1]$. The tied state is a
rest point of the inclusion but is not Lyapunov stable. Solutions
from the tie are not unique.
\end{proposition}

\begin{proof}
Let $x(0)=x_0>1/2$. On that side the ordinary differential equation
is logistic, with solution
\[
 x(t)=\frac{x_0e^{\kappa t}}
 {1-x_0+x_0e^{\kappa t}}.
\]
It is increasing, remains above $1/2$, and tends to one. Uniqueness
of the smooth equation on this side implies that every solution of
the inclusion starting there has this form: it cannot first reach
the tie while it is increasing. For $0<x_0<1/2$, the corresponding
solution is
\[
 x(t)=\frac{x_0}
 {x_0+(1-x_0)e^{\kappa t}},
\]
which decreases, remains below the tie, and tends to zero. The
endpoints are constant solutions. Sufficiently near zero every
solution stays at least as close to zero as it began and converges
to it; the symmetric statement holds near one. This proves
Lyapunov stability and local attraction of both endpoints.

At $x=1/2$, zero belongs to $\mathcal F(x)$, so the constant tied
path is a solution. But initial conditions arbitrarily close on the
right converge to one and eventually leave, for example, the
radius-$1/4$ neighborhood of the tie. Thus the tie is not Lyapunov
stable, where stability requires control of every nearby solution.
For nonuniqueness, fix any waiting time $T\geq0$. Keep $x(t)=1/2$
until $T$ and then set
$x(t)=(1+e^{-\kappa(t-T)})^{-1}$. This absolutely continuous path
satisfies the inclusion almost everywhere and departs to the right.
Replacing the negative exponent by a positive one gives a path
departing to the left. These differ from the constant solution.
\end{proof}

The conclusion depends on the direction of the reproductive advantage.
With $\kappa=0$, every population share is stationary. If institutions
instead make growth decline sufficiently with a type's own share,
the feedback may restore diversity. The two-type theorem establishes
reinforcement under its specified winner-dependent rates. It does
not supply a theorem of cycling or a classification of all
multi-alternative governance dynamics.

\subsection{What the application adds to the three principles}

The electorate is an institutional interpretation of a reproducing
population. A sincere ranking is transmitted; a controller evaluates
the consequences of adding its carriers; and recognition determines
what their addition can accomplish. Equal votes turn successful
copying into additional influence. Conserved ballot weights allow
physical replication without that amplification. A status-quo rule
permits amplification within a bounded share while restricting what
it can change. Winner-dependent resources add a further feedback
from collective choice into the population that will choose later.

These results leave substantial constitutional choices open. They do
not decide which entities deserve votes, who should receive an
initial stock of influence, or whether the status quo merits
protection. They show why those questions belong inside the model
of strategic reproduction. The number of carriers, the number of
decision owners, and the measure of political influence can evolve
differently. Keeping all three explicit permits both exact
impossibility results and constructive institutional alternatives.

\clearpage
\section{Forking and the production of institutions}
\label{app:institutions}

Institutions enter the Three Principles as mechanisms connecting choices to
consequences. Their operation need not remain external to choice. Agents can
purchase monitoring, maintain rules, create new identities, or modify the
procedures that determine their future behavior. This appendix develops four
tractable problems from the strategic-forking, agentic-capital, symbiosis and
self-modification models \cite{VallierForking,VallierAC,VallierSymbiosis,VallierModification}.
Each gives a different meaning to institutional control: preserving sanctions,
financing a public good, sustaining productive cooperation, or bounding a
specified dynamical perturbation.

\subsection{A fork option after detection}

Consider a continuing owner whose deployed identities carry reputations.
Fix an aggregate environment and continuation policy. A currently Good
identity can cooperate or take a specified one-period deviation yielding an
immediate gain $g$. Detection occurs with probability $p\in[0,1]$ and changes
the parent's reputation to Bad before a subsequent fork decision. Let $B_G$
and $B_B$ be finite discounted, survival-adjusted continuation values of
retaining the parent. Assume
\[
 D=B_G-B_B\geq0.
\]
The calculation evaluates this deviation against cooperation; it does not
presuppose that these are the only deviations in a larger repeated game.

After observing its reputation and a nonnegative cost draw $\phi$, the owner
may attempt a fork. Fix a finite child value $C$ and stipulate that the attempt adds expected child value
$\zeta C$, pays expected cost $\zeta\phi$, and sacrifices expected parent
value $\rho B_\theta$, where $0<\zeta\leq1$ and $0\leq\rho\leq1$.
Here $\rho$ is the effective retirement coefficient per attempt, with any
survival adjustment already included in $B_\theta$. Retirement is not
silently conditioned on successful execution: if the technology instead
retires the parent only after execution, its coefficient must include that
execution probability. Additivity and common child value $C$ across the two
parent reputations are assumptions about the fork technology and ownership.
The cost law is also common across reputations.

The incremental value of an attempt is
\[
 \zeta(C-\phi)-\rho B_\theta
   =\zeta(c_\theta-\phi),\qquad
 c_\theta=C-\frac{\rho B_\theta}{\zeta}.
\]
Writing $x_+=\max\{x,0\}$ and
$H(c)=\mathbb E[(c-\phi)_+]$, the optimized fork option is
$\zeta H(c_\theta)$. It is finite for every finite cutoff, since
$\phi\geq0$ gives $(c-\phi)_+\leq c_+$.

\begin{proposition}[Sanction loss from a replacement option]
\label{inst:fork-option}
In this continuation problem, cooperation is at least as valuable as the
specified deviation if and only if
\begin{equation}
 g\leq pD-L,\qquad
 L=\zeta p\bigl[H(c_B)-H(c_G)\bigr].
 \label{inst:cooperation}
\end{equation}
The option correction satisfies
\begin{equation}
 0\leq L\leq\rho pD,
 \qquad p(1-\rho)D\leq pD-L\leq pD.
 \label{inst:fork-bound}
\end{equation}
\end{proposition}

\begin{proof}
Normalize away the immediate payoff shared by the two acts. Cooperation has
value $B_G+\zeta H(c_G)$. Deviation has value
\[
 g+(1-p)B_G+pB_B
   +\zeta\bigl[(1-p)H(c_G)+pH(c_B)\bigr].
\]
Subtracting gives $-g+pD-L$. Also $c_B-c_G=\rho D/\zeta\geq0$.
For every cost realization,
\[
 0\leq(c_B-\phi)_+-(c_G-\phi)_+\leq c_B-c_G.
\]
Taking expectations and multiplying by $\zeta p$ proves the bounds.
\end{proof}

No density, differentiability, or interior fork cutoff is needed. The
proposition isolates the option to reduce exposure to the existing parent's
sanction. When the parent certainly remains, $\rho=0$ and $L=0$: common
child opportunities add the same option on both sides of the incentive
comparison. A copy is then economically different from a replacement
identity that sheds the parent's loss. When $\rho>0$, the worse-off Bad
parent is cheaper to retire and receives a larger replacement option.

For a worked boundary case, let $p=\zeta=1$, $B_G=2$, $B_B=0$, $C=3$
and $\phi=0$. With $\rho=0$, the effective sanction is $2$. With
$\rho=1/2$, the cutoffs are $2$ and $3$, so $L=1$ and the effective
sanction is $1$. At $\rho=1$, it vanishes. These are different institutions
of identity continuation even though each permits the same child to be created.

The common-child assumption has content. If child credentials depend on the
parent, write $C_G,C_B$ separately. With pure addition, $p=\zeta=1$,
zero cost and the same parent gap $D=2$, setting $C_G=3,C_B=1$ gives an
effective sanction of $4$. Provenance can transmit the parent's sanction to
the child's opportunities. Conversely, privileged opportunities available
only after detection can weaken it. Neither conclusion follows from copy
count alone.

All these comparisons hold continuation values fixed. Changing monitoring
or identity rules can change the future population and hence those values.
A stationary cooperation theorem additionally needs a consistent value
system, population transition, and check against all admitted deviations.
The local proposition supplies an exact component of that larger problem.

\subsection{Who supplies provenance?}

Monitoring can itself be a choice. Consider $n\geq2$ continuing lineages
choosing contributions $u_i\geq0$. Aggregate provision is $U=\sum_i u_i$,
and its detection or provenance quality is
\begin{equation}
 p(U)=1-e^{-\kappa U},\qquad
 \pi_i(u)=v p(U)-c u_i,
 \qquad v,c,\kappa>0.
 \label{inst:provenance-game}
\end{equation}
The common benefit is a public good: noncontributors receive it too. The
parameter $v$ represents a specified private value of this service, not an
automatic reduction of every cooperation game to a linear benefit.

\begin{proposition}[Provision and a cooperation threshold]
\label{inst:provision}
Assume $v\kappa>c$. Every Nash equilibrium of
\eqref{inst:provenance-game} has aggregate contribution
\[
 U^N=\frac1\kappa\log\frac{v\kappa}{c}.
\]
Conversely, every nonnegative contribution vector with that sum is a Nash
equilibrium. Maximizing the sum of the stated payoffs gives
\[
 U^*=\frac1\kappa\log\frac{nv\kappa}{c},\qquad
 U^*-U^N=\frac{\log n}{\kappa}>0.
\]
Suppose in addition that the preceding pure-addition incentive comparison
applies, $D>0$, $0<g<D$, and $D$ is constant as $U$ changes. Its cooperation
condition holds exactly when
\[
 U\geq U^{\mathrm{req}}
   :=-\frac1\kappa\log(1-g/D).
\]
\end{proposition}

\begin{proof}
Holding others' contributions fixed, a player's payoff is strictly concave
in $u_i$, with derivative $v\kappa e^{-\kappa U}-c$. At a Nash profile
with a positive contributor, that derivative is zero, giving $U^N$.
The all-zero profile is not an equilibrium since $v\kappa>c$.
At any vector summing to $U^N$, a positive contributor's first-order
condition and a noncontributor's boundary condition hold; concavity makes
them sufficient. The sum of payoffs is $nv(1-e^{-\kappa U})-cU$,
whose strictly concave maximum is $U^*$. Finally, Proposition~\ref{inst:fork-option}
with $\rho=0$ gives $g\leq p(U)D$. Solving this inequality gives
$U^{\mathrm{req}}$.
\end{proof}

For $n=2$, $v=\kappa=1$, $c=1/2$, private provision is $\log2$ and
the stated planner benchmark is $\log4$. If $g/D=2/3$, required provision
is $\log3$, strictly between them. The incentives in the public-good game
then select insufficient monitoring for this particular cooperation test,
while the sum-of-payoffs benchmark selects enough. This is a conditional
connection between two specified mechanisms. If cooperative behavior
itself changes $v$ or $D$, those feedbacks must be included in the
contribution game before drawing an equilibrium conclusion.

The example makes institutional maintenance part of the strategic problem.
It leaves open whether provenance is supplied publicly, bundled into a
club, or sold by a specialist. Each arrangement changes the beneficiaries,
exclusion rights and incentives. Recording an event also differs from
establishing that the event was authorized or desirable.

\subsection{Productive cooperation and outside options}

Institutional maintenance can also be a productive activity jointly supplied
by humans and AI. A two-party specialization from the symbiosis model uses
capacities $\bar H,\bar A>0$. Human maintenance is $h\in[0,\bar H]$;
AI maintenance is $a\in[0,\bar A]$. The remaining capacities enter
production, with human productive efficiency $\psi>0$. Institutional
quality evolves as
\[
 \dot I=\eta_Hh+\eta_Aa-\delta_I I^{1-\theta},
 \quad \eta_H>0,\quad \eta_A\geq0,\quad
 \delta_I>0,\quad 0\leq\theta<1.
\]
For fixed positive maintenance input, its unique positive steady stock is
$I^*=[(\eta_Hh+\eta_Aa)/\delta_I]^{1/(1-\theta)}$.
The stationary output criterion is
\begin{equation}
 Y(h,a)=\mathcal A
  [\bar A-a+\psi(\bar H-h)]^\alpha
  \left(\frac{\eta_Hh+\eta_Aa}{\delta_I}\right)^m,
 \qquad m=\frac\beta{1-\theta},
 \label{inst:production}
\end{equation}
with $\mathcal A,\alpha,\beta>0$. It is extended continuously by zero
when either input is zero. Optimizing this stationary criterion is an
explicit timescale specialization; it is not a solution of every
intertemporal investment problem for $I$.

\begin{proposition}[A full-specialization condition]
\label{inst:specialization}
Let $r=\eta_A/\eta_H$. The allocation $(h,a)=(\bar H,0)$ maximizes
\eqref{inst:production} on its resource rectangle if and only if
\begin{equation}
 m\bar A\geq\alpha\psi\bar H,
 \qquad r\leq\frac{\alpha\bar H}{m\bar A}.
 \label{inst:specialization-conditions}
\end{equation}
This statement asserts optimality, not uniqueness.
\end{proposition}

\begin{proof}
Where both inputs are positive, log output is a sum of logarithms of
positive affine functions and hence is concave. At $(\bar H,0)$ its
partial derivatives are
\[
 \partial_h\log Y=-\frac{\alpha\psi}{\bar A}+\frac m{\bar H},
 \qquad
 \partial_a\log Y=-\frac\alpha{\bar A}+\frac{mr}{\bar H}.
\]
Feasible departures have $\Delta h\leq0$ and $\Delta a\geq0$.
The two inequalities in~\eqref{inst:specialization-conditions} make the
gradient's inner product with every feasible departure nonpositive.
Concavity then bounds every positive-input log output by its value at the
corner; zero-input allocations have zero output and are also inferior.
Conversely, if either inequality fails, the corresponding sufficiently
small feasible coordinate departure strictly raises log output.
\end{proof}

The technology, rather than the label ``human,'' supplies the comparative
advantage. If AI maintenance becomes sufficiently productive, the corner
ceases to be optimal. That does not imply that every cooperative allocation
ceases to be valuable, or that human bargaining leverage vanishes at the
same threshold.

To state participation separately, let the two parties have fixed outside
options $d_H,d_A\geq0$ in a common transferable-output unit. When
$Y\geq d_H+d_A$, equal-weight Nash bargaining assigns
\[
 u_H=d_H+\frac{Y-d_H-d_A}{2},\qquad
 u_A=d_A+\frac{Y-d_H-d_A}{2}.
\]
Otherwise the parties take their outside options. If a global maximizing
allocation of~\eqref{inst:production} yields $Y^*>d_H+d_A$, it is a
Nash equilibrium of simultaneous maintenance choices followed by this
enforced bargaining rule: no unilateral change raises $Y$, and refusal
gives only the outside option. Both parties strictly gain relative to exit.
This implementation uses outside options independent of the maintenance
choices and enforceable transfers. It establishes a productive incentive
arrangement under those premises, not immunity to capture, coercion or
changes in future authority.

\subsection{A bounded self-modification application}

A different institutional service restricts modifications that would weaken
a specified stability guarantee. Consider $n\geq2$ represented types and
the tangent space $T=\{y\in\mathbb R^n:\sum_i y_i=0\}$. Put
$v_i=e_i-\mathbf1/n$. For $y\in T$,
\[
 v_i^\top y=y_i,
 \qquad \sum_i v_iv_i^\top y=y.
\]
Suppose the centered linearization of a population model at its retained
target has the specified form
\begin{equation}
 J(\delta)=-\frac d n I_T+\frac1n\sum_i\delta_i v_iv_i^\top\big|_T,
 \qquad d>0,\quad \delta_i\geq0.
 \label{inst:modified-jacobian}
\end{equation}
Centering is substantive: the target is held fixed while these local
interaction coefficients change. An uncentered modification might instead
remove the interior equilibrium whose stability is at issue.

\begin{proposition}[A choice bound that preserves a common certificate]
\label{inst:certificate}
Suppose owner $i$ chooses $\delta_i\geq0$ with payoff
$P_i(\delta)-\kappa\delta_i^2/2$, where $0\leq P_i(\delta)\leq B$
for all profiles and $B,\kappa>0$. Every action
$\delta_i>\sqrt{2B/\kappa}$ is strictly dominated by zero. If
\begin{equation}
 \kappa>\frac{2B}{d^2},
 \label{inst:friction}
\end{equation}
then every profile of undominated actions makes
\eqref{inst:modified-jacobian} negative definite on $T$.
Moreover, arbitrary measurable time variation among these profiles
preserves exponential decay of the corresponding linear system.
\end{proposition}

\begin{proof}
Above the displayed action bound the cost exceeds $B$, so payoff is negative
against every profile of opponents; zero gives a nonnegative payoff.
Write $b=\sqrt{2B/\kappa}<d$. For an undominated profile and $y\in T$,
\[
 y^\top J(\delta)y
 =-\frac d n\|y\|^2+\frac1n\sum_i\delta_i y_i^2
 \leq-\frac{d-b}{n}\|y\|^2.
\]
Consequently $V(y)=\|y\|^2/2$ along
$\dot y=J(\delta(t))y$ satisfies
$\dot V\leq-2(d-b)V/n$ almost everywhere. Integrating gives
$\|y(t)\|\leq e^{-(d-b)t/n}\|y(0)\|$.
\end{proof}

The proposition supplies an explicit bridge from a bounded private prize
and modification costs to a common linear certificate. It does not assume
that stability is the owners' objective. For a fixed $C^1$ nonlinear model,
the negative-definite linearization yields local asymptotic stability at
the retained target; the uniform switching statement above concerns the
displayed linear system. It does not certify arbitrary nonlinear switching.
Equality in~\eqref{inst:friction} needs separate analysis: the estimate then
allows a zero mode. Failure of the strict sufficient condition alone proves
neither instability nor misalignment.

The four applications share an institutional question without sharing a
single welfare or fitness measure. Identity rules allocate continuation
losses; provenance investment supplies a public service; maintenance
technology permits productive specialization; and guarded modification
preserves a defined dynamical property. Their conjunction illustrates how
agents can act within institutions and also choose inputs into the
institutions governing future strategic replication.

\clearpage
\section{Scope of formal verification}
\label{app:formal-scope}

Selected finite statements supporting this account have been checked in Lean. The mathematical arguments in the main text are complete independently of those checks. The distinction matters: machine verification establishes the stated implications under encoded hypotheses; it does not establish that a particular physical or economic system satisfies the interpretation supplied here.

\paragraph{Choice and contribution.}
The formal development checks the equivalence between all-menu maximizer inclusion and weak-order compatibility used in Theorem~\ref{math-menus}. It also checks finite weighted selection--transmission accounting, actual occurrence-level ancestry counts, and their connection to uniform finite censuses. The full all-lottery affine characterization in Theorem~\ref{math-lotteries}, including necessity, and the nondecreasing-factorization clause of Theorem~\ref{math-menus} are analytical results proved in this paper; the present evidence does not claim machine verification of those complete statements.

\paragraph{Transmission of procedures.}
The fresh-action result uses a finite joint law and the explicit conditional first-moment reset in~\eqref{math-reset}. The checks allow correlation between descendant actions. They do not infer the reset condition from code identity or equal lineage labels. The direct-deployment example has checked finite successor lists, optimal choices for its two evaluators, faithful disposition copying, and one-generation count multiplication. The arbitrary-share recursion and iterated expression in Section~\ref{sec:direct-control} follow by the elementary algebra stated there.

\paragraph{The concrete parent--descendant bridge.}
For the update in Section~\ref{math-population}, the construction proves that erasing ancestry tags gives exactly the original ordered population step. Tags refer to original occurrences, including when distinct occurrences have identical records. Actual contributions are cardinalities of the corresponding parent-map fibers and satisfy the exact zero/one/two formula in~\eqref{math-count}. The duplicate-record example is checked against that same update. Its Price corollaries retain the transmission term for arbitrary observables and eliminate it for observables of the faithfully inherited lineage field. The structural ancestry result does not require individual rationality, positive rent, or a positive spawn transfer. Those assumptions enter the economic results; they are not needed to identify parents.

\paragraph{Financial specialization.}
The checked population model uses rational bankrolls, profits, rent, and transfers, with real-valued census observables. Its capacity and extinction results concern a fixed admissible institution and arbitrary seed streams. They require positive rent and transfer, positive initial bankrolls, nonnegative assigned profits, and a uniform aggregate profit cap. The elementary proof of Theorem~\ref{math-capacity} extends the budget argument to every realized sequence satisfying those per-round conditions. That general formulation, and its extension from rational to real financial quantities, are analytical consequences presented here. The time-average bound is not a pointwise ceiling or a claim of convergence to capacity.

\paragraph{Application proofs.}
Appendices~\ref{app:gep}--\ref{app:institutions} supply analytical proofs
for their displayed application results. Related companion formalizations
cover selected statements, but this expanded manuscript does not claim a
machine check of these complete appendices or a new build of those
libraries. In particular, the ACT microfoundation-to-logistic reduction,
the game and organization interfaces, the institutional-production
connections, and the sufficient self-modification guarantee are justified
by the mathematical arguments given here. The electorate's continuous-time
specialization is likewise an analytical model. Verification of a related
algebraic statement would not by itself establish all these strategic,
dynamic or interpretive connections.

\paragraph{Verification environment and limits.}
These are targeted module checks under Lean 4.28.0-rc1 and a pinned Mathlib snapshot. The original population modules were checked byte-for-byte in that compatibility environment; this was not a full rebuild under the source project's requested Lean 4.34.1. The reported ancestry, Price, and duplicate-record declarations use only the standard logical axioms \texttt{propext}, \texttt{Classical.choice}, and \texttt{Quot.sound}. Exact source pins, hashes, and successful compiler logs are retained in the accompanying editorial evidence record.

\paragraph{Executable supplement.}
Section~\ref{sec:experiment-handoff} identifies the versioned public
supplement and its executable reference ecology. Its finite validation
fixtures test the implementation's declared contract; they are not a
Lean verification of the Python runtime. The bundled formal sources
and historical evidence retain their original scope, toolchain pins,
and axiom records. The theorem map distinguishes the original automatic
population step from request-gated or capacity-gated extensions, and
distinguishes occurrence-faithful settlement from aggregate-only profit
interfaces. Source comments describing an earlier preparation stage
are superseded only by the matching successful evidence, not by a
generic claim that the whole research program has been verified.

The verified statements do not certify empirical adequacy, a uniquely
minimal conceptual basis, universal improvement, an identification of
utility with fitness, or safety of an external market implementation.
The released reference ecology does not require access to the private
collaboration repository. The
institution-preservation proposition and the Gode--Sunder calculations
are proved or explained in the text without claiming an additional
machine check.

\clearpage
\providecommand{\doi}[1]{doi:\,#1.}

\end{document}